\documentclass[11pt,a4paper]{article}
\usepackage{amsmath, amsfonts, amssymb}
\usepackage{a4wide}
\usepackage{parskip}
\usepackage{enumitem}
\usepackage{xcolor}

\usepackage{hyperref}
\usepackage{booktabs}
\usepackage{caption}
\usepackage{siunitx}
\usepackage{tabularx}
\usepackage{graphicx}
\usepackage{adjustbox}
\usepackage{multirow}
\usepackage{amsthm}
\usepackage{slashbox}
\usepackage{diagbox}
\usepackage{tikz}
\usetikzlibrary{matrix}
\usepackage[noadjust]{cite}
\def\proof{\noindent{\textit{Proof. }}}
\def\qed{\hfill {$\square$}\goodbreak \medskip}

\newtheorem{theorem}{Theorem}[section]
\newtheorem{lemma}[theorem]{Lemma}
\theoremstyle{definition}
\newtheorem{definition}[theorem]{Definition}
\newtheorem{example}[theorem]{Example}
\theoremstyle{conjecture}

\theoremstyle{proposition}
\newtheorem{proposition}[theorem]{Proposition}
\theoremstyle{remark}
\newtheorem{remark}[theorem]{Remark}

\theoremstyle{corollary}

\numberwithin{equation}{section}

\usepackage{tikz,xcolor,hyperref}

\definecolor{lime}{HTML}{A6CE39}
\DeclareRobustCommand{\orcidicon}{%
	\begin{tikzpicture}
		\draw[lime, fill=lime] (0,0) 
		circle [radius=0.16] 
		node[white] {{\fontfamily{qag}\selectfont \tiny ID}};
		\draw[white, fill=white] (-0.0625,0.095) 
		circle [radius=0.007];
	\end{tikzpicture}
	\hspace{-2mm}
}

\foreach \x in {A, ..., Z}{%
	\expandafter\xdef\csname orcid\x\endcsname{\noexpand\href{https://orcid.org/\csname orcidauthor\x\endcsname}{\noexpand\orcidicon}}
}

\begin{document}
	\date{}
	{\vspace{0.01in}
		\title{
		Certain binary minimal codes constructed using simplicial complexes}
		\author{{\bf Vidya Sagar\footnote{email: {\tt vsagariitd@gmail.com}}\orcidA{}  \;and  \bf Ritumoni Sarma\footnote{email: {\tt ritumoni407@gmail.com}}\orcidB{}} \\ Department of Mathematics,\\ Indian Institute of Technology Delhi,\\Hauz Khas, New Delhi-110016, India. }
		\maketitle
		\begin{abstract}
			In this manuscript, we work over the non-chain ring $\mathcal{R} = \mathbb{F}_2[u]/\langle u^3 - u\rangle $. Let $m\in \mathbb{N}$ and let $L, M, N \subseteq [m]:=\{1, 2, \dots, m\}$. For $X\subseteq [m]$, define $\Delta_X:=\{v \in \mathbb{F}_2^m : \textnormal{Supp}(v)\subseteq X\}$ and $D:= (1+u^2)D_1 + u^2D_2 + (u+u^2)D_3$, an ordered finite multiset consisting of elements from $\mathcal{R}^m$, where $D_1\in \{\Delta_L, \Delta_L^c\}, D_2\in \{\Delta_M, \Delta_M^c\}, D_3\in \{\Delta_N, \Delta_N^c\}$. The linear code $C_D$ over $\mathcal{R}$ defined by $\{\big(v\cdot d\big)_{d\in D} : v \in \mathcal{R}^m \}$ is studied for each $D$. Further, we also consider simplicial complexes with two maximal elements in the above work. We study their binary Gray images and the binary subfield-like codes corresponding to a certain $\mathbb{F}_{2}$-functional of $\mathcal{R}$. Sufficient conditions for these binary linear codes to be minimal and self-orthogonal are obtained in each case. Besides, we produce an infinite family of optimal codes with respect to the Griesmer bound. Most of the codes obtained in this manuscript are few-weight codes.
			\medskip
			
			
			\medskip
			
			
		\end{abstract}
		
		\section{INTRODUCTION}
		This manuscript studies the linear code $C_D = \{(v\cdot d)_{d\in D} : v\in \mathcal{R}^m\}$ over $\mathcal{R}=\mathbb{F}_2[u]/\langle u^3 - u\rangle$, where $D \subseteq \mathcal{R}^m$ is an ordered finite multiset. Ding et al. in \cite{Ding} first defined the above construction of $C_D$ in order to generalize the Mattson-Solomon transform for cyclic codes. And due to this construction of linear codes, simplicial complex got huge attention in algebraic coding theory. By using simplicial complexes, several interesting linear codes have been constructed in the recent past. The authors in \cite{Hyun_Lee} studied linear codes by selecting certain defining sets derived using simplicial complexes and obtained a class of binary optimal linear codes. Zhu et al. in \cite{Zhu_Wei} studied linear codes over $\mathbb{F}_4$ and obtained minimal optimal codes. Yansheng et al. in \cite{Wu_Li} studied linear codes over $\mathbb{F}_{4}$ and produced two infinite classes of optimal codes; motivated by this work, the authors in \cite{Sagar_Sarma} constructed linear codes over $\mathbb{F}_8$ using simplicial complexes and produced three infinite families of optimal codes. They also discussed sufficient conditions for minimality of these codes by using Ashikhmin-Barg condition (see \cite{suffConMinimalcode}). Recently, the authors in \cite{GeneralCase} generalized the work of \cite{Sagar_Sarma} for finite fields of characteristic $2$ by using LFSR sequences. The weight distribution (see \cite{wchuffman}) of a linear code contains crucial information regarding error-detecting as well as error-correcting capacity of the code, and it allows the computation of the error probability of error-detection and -correction with respect to some algorithms \cite{Klove}. Few-weight codes are useful because of their connection with strongly regular graphs, difference sets and finite geometry \cite{Few1, Few2}. The minimum Hamming distance of linear codes are well known for their importance in determining error-correcting capacity. Due to this reason, finding optimal linear codes has become one of the central topics for researchers. In \cite{Hyun_Lee}, the authors showed how optimal codes can be utilized for the construction of secret sharing schemes with nice access structures following the framework discussed in \cite{YuanDing}.\par 
		The study of minimal codes is an active research topic nowadays because of its applications in secure two-party computation and secret sharing schemes \cite{CarletDing, Chabanne_two_party, ShamirSSS, YuanDing}. This special class of codes is also important as these can be decoded by using the minimum distance decoding rule \cite{suffConMinimalcode}. Normally, it is difficult to identify all the minimal codewords of a given code even over a finite field of characteristic $2$. Many researchers have studied linear codes over a single or mixed alphabet using simplicial complexes and produced minimal and optimal codes (see \cite{Hyun_Kim, Shi_guan, NewAdd, shi_xuan, shi_x, shi_qian,shi_nonchain,  shi_x2,   mixed1, mixed2, wu_zhu}) in the recent past.\par
		Motivated by the research in \cite{shi_nonchain}, a natural endeavor is to explore the construction of linear codes over other finite rings using simplicial complexes and obtain minimal and optimal codes. We, in this manuscript, construct linear codes over $\mathcal{R}=\mathbb{F}_2[u]/\langle u^3 - u\rangle $ whose defining sets are obtained from certain simplicial complexes generated by one and two maximal elements, and find their Lee-weight distributions. We study their binary Gray images and their binary subfield-like codes corresponding to certain $\mathbb{F}_{2}$-functional of $\mathcal{R}$. An infinite family of optimal codes are produced with respect to the Griesmer bound. Sufficient conditions for these binary codes to be minimal and self-orthogonal are established. We obtain the parameters and the weight distributions of the 1st order Reed-Muller codes of length $2^{m-1}$, where $m>1$, as the parameters and the weight distributions of certain subfield-like codes. We give a table that shows that our codes include several binary optimal codes with new parameters.\par
		The remaining sections of this manuscript are arranged as follows. Preliminaries are presented in the next section. Section \ref{section3} studies the construction of linear codes using simplicial complexes generated by one and two maximal elements. In Section \ref{section4}, we obtain Gray images of linear codes constructed using simplicial complexes generated by a single maximal element in the previous section and find their minimality and self-orthogonality conditions. Section \ref{section5} studies certain binary subfield-like codes from codes considered in Section \ref{section3} and finds conditions for the optimality, minimality, and self-orthogonality of these codes. In Section \ref{section6}, we compare our binary codes with the recent research in this direction. In this section, we also present a table that shows that our codes include several optimal codes with new parameters. Section \ref{section7} concludes this manuscript.
		\section{Definitions and Preliminaries}\label{section2}
		Throughout the manuscript, $\mathcal{R}=\mathbb{F}_{2}[u]/\langle u^{3}-u \rangle$, where $\mathbb{F}_{2}$ is the binary field. Observe that $\mathcal{R}$ is a non-chain principal ideal ring. The group of units is $U(\mathcal{R})=\{1, 1+u+u^2\}$ and the set of non-units is $\mathcal{R}\setminus U(\mathcal{R})=\{0, u, u^2, u+u^2, 1+u, 1+u^2\}$.
	    \begin{lemma}\cite{RingR}
	    	$\mathcal{R}$ can be written as $(1+u^2)\mathbb{F}_2+u^2\mathbb{F}_2+(u+u^2)\mathbb{F}_2$ uniquely.	    	
	    \end{lemma}	    
		Let $\Phi : \mathcal{R} \longrightarrow \mathbb{F}_2^3$ be the \textit{Gray map} given by 
		\begin{equation}\label{PhiEq}
			\Phi(a+ub +u^2d) = (a+b, b+d,d).
		\end{equation}
		This extends to $\Phi : \mathcal{R}^m \longrightarrow \mathbb{F}_2^{3m}$ where $\Phi(r+us+u^2t)=(r+s, s+t, t)$. 
		\begin{definition}
			An $\mathcal{R}$-submodule $C$ of $\mathcal{R}^m$ is a \textit{linear code} over $\mathcal{R}$ and $m$ is its length.
		\end{definition} 
		Let $v, w \in \mathbb{F}_2^m$. Then the \textit{Hamming weight} of $v$, denoted by $wt_{H}( v )$, is the number of nonzero entries in $v$. The \textit{Hamming distance} between $v$ and $w$ is $d_H(v, w) = wt_{H}(v-w)$.\\
		Let $x=\alpha +u\beta +u^2\gamma$, $y=\alpha' +u\beta' +u^2\gamma' \in \mathcal{R}^m$, where $\alpha, \beta, \gamma, \alpha', \beta', \gamma'\in \mathbb{F}_2^m$. Then the \textit{Lee weight} of $x$ is $wt_{Lee}(x) = wt_{H}(\Phi(x)) = wt_{H}(\alpha + \beta) + wt_{H}(\beta +\gamma ) + wt_H(\gamma)$. The \textit{Lee distance} between $x$ and $y$ is $d_{L}(x, y) = wt_{Lee}(x-y)$. Thus, the map $\Phi$ is an isometry. Note that the image of a linear code over $\mathcal{R}$ under the Gray map $\Phi$ is a binary linear code. Suppose $C$ is a linear code over $\mathcal{R}$ of length $n$, where $n \in \mathbb{N}$. Let $A_i$ be the cardinality of the set that contains all codewords of $C$ having Lee weight $i$, $0\leq i\leq 3n$. Then the homogeneous polynomial in two variables 
		\begin{equation*}
			Lee_C(X, Y)=\sum_{c\in C}X^{3n-wt_{Lee}(c)}Y^{wt_{Lee}(c)}
		\end{equation*} is called the \textit{Lee weight enumerator} of $C$ and the string $(1, A_1,\dots ,A_{3n})$ is called the \textit{Lee weight distribution} of $C$. Similarly, we can define the Hamming weight enumerator and the Hamming weight distribution for a linear code over a finite field. In addition, if the total number of $i\geq 1$ such that $A_i\neq 0$ is $l$, then $C$ is called an $l$-\textit{weight linear code}. Every $1$-weight linear code is an equidistant linear code. In \cite{Bonisoli}, Bonisoli characterized all equidistant linear codes over finite fields.
	    \begin{theorem}\cite{Bonisoli}\textnormal{(Bonisoli)}\label{Bonisoli}
	    Suppose $C$ is an equidistant linear code over $\mathbb{F}_q$. Then $C$ is equivalent to the $r$-fold replication of a simplex code, possibly with added $0$-coordinates.
        \end{theorem}
        An $[n, k, d]$-linear code $C$ is called \textit{distance optimal} if there exists no $[n,k,d+1]$-linear code (see \cite{wchuffman}). Next we recall the Griesmer bound.
        \begin{lemma}\label{griesmerbound}\cite{Griesmer}
        \textnormal{(Griesmer Bound)} If $C$ is an $[n,k,d]$-linear code over $\mathbb{F}_q$, then we have
        \begin{equation}\label{GriesmerBound}
	     \sum\limits_{i=0}^{k-1}\left\lceil \frac{d}{q^i}\right\rceil \leq n,
        \end{equation}  	
         where $\lceil \cdot \rceil$ denotes the ceiling function.
        \end{lemma}
        A linear code is called a \textit{Griesmer code} if equality holds in Equation \eqref{GriesmerBound}. Note that every Griesmer code is distance optimal, but the converse needs not be true.
	
	
	\section{Construction of linear codes over the ring $\mathcal{R}$}\label{section3}
	For $m\in \mathbb{N}$, we shall write $[m]$ to denote the set $\{1, 2,\dots ,m\}$.
	For any $n, m\in \mathbb{N}$, let $D=\{d_1<d_2<\dots <d_n\} \subseteq \mathcal{R}^m$ be an ordered multiset of cardinality $n$. Define \cite{wchuffman},
	\begin{equation}\label{genericConstruction}
		C_{D}=\{(v\cdot d_1, v\cdot d_2, \dots , v\cdot d_n): v\in \mathcal{R}^m\}   , 	
	\end{equation}
	where $x\cdot y=\sum\limits_{i=1}^{m}x_iy_i$ for $x,y\in \mathcal{R}^m$. Then $C_D$ forms a linear code of length $\vert D\vert $ over $\mathcal{R}$.	
	The ordered set $D$ is called the \textit{defining set} of $C_{D}$. Note that on changing the ordering of $D$ we will get a code which is permutation equivalent to $C_D$. If $D$ is chosen appropriately, $C_{D}$ may possess good parameters. For $w\in \mathbb{F}_{2}^m$, the set $\textnormal{Supp}(w)=\{ i\in [m]: w_i = 1\}$ is called the \textit{support} of $w$. Note that the Hamming weight of $w\in \mathbb{F}_2^m$ is $wt_H(w)=\vert \textnormal{Supp}(w)\vert $. For $v,w\in \mathbb{F}_{2}^m$, one says that $v$ covers $w$ if $\textnormal{Supp}(w)\subseteq \textnormal{Supp}(v)$. If $v$ covers $w$, we write $w\preceq v$.\\
	Consider the map $\psi: \mathbb{F}_2^m\longrightarrow 2^{[m]}$ is defined as $\psi(w)=\textnormal{Supp}(w)$, where $2^{[m]}$ denotes the power set of $[m]$. Note that $\psi$ is a bijective map. Now onwards, we will write $w$ instead of Supp($w$) whenever we require.
	\begin{definition}
		A subset $\Delta$ of $\mathbb{F}_{2}^m$ is called a \textit{simplicial complex} if $v\in \Delta, w\in \mathbb{F}_2^m$ and $w\preceq  v$ $\implies$ $w\in \Delta$. An element $v\in \Delta$ is called a \textit{maximal element of} $\Delta$ if for $w\in \Delta$, $v\preceq w$ $\implies$ $v=w$. A simplicial complex can have more than one maximal elements.
	\end{definition}
	
	\subsection{Using simplicial complexes having one maximal element}
	In this subsection, we study linear codes over $\mathcal{R}$ using simplicial complexes generated by one maximal element.\\
	Let $L\subseteq [m]$. The \textit{simplicial complex generated by} $L$ is denoted by $\Delta_{L}$ and is defined as 
	\begin{equation}
		\Delta_{L}=\{w\in \mathbb{F}_{2}^m \vert \textnormal{ Supp}(w)\subseteq L\}.
	\end{equation}
	Note that $\psi^{-1}(L)$ is the only maximal element of $\Delta_L$ and $\vert \Delta_{L}\vert =\vert 2^L\vert=2^{\vert L\vert}$. \par
	Given a subset $P$ of $\mathbb{F}_{2}^m$, define the polynomial (referred to as an \textit{m-variable generating function}, see \cite{Chang}) $\mathcal{H}_{P}(y_1, y_2,\dots , y_m)$ by
	\begin{equation}
		\mathcal{H}_P(y_1,y_2,\dots ,y_m)=\sum\limits_{v\in P}\prod_{i=1}^{m}y_i^{v_i}\in \mathbb{Z}[y_1,y_2,\dots ,y_m],
	\end{equation}
	where $v=(v_1,\dots ,v_m)$ and $\mathbb{Z}$ denotes the ring of integers.\\
	We recall a lemma from \cite{Chang}.
	\begin{lemma}\cite{Chang}\label{generatinglemma}
		Suppose $\Delta\subseteq \mathbb{F}_{2}^m$ is a simplicial complex and $\mathcal{F}$ consists of its maximal elements. Then 
		\begin{equation}
			\mathcal{H}_{\Delta}(y_1,y_2,\dots ,y_m)=\sum\limits_{\emptyset\neq S\subseteq\mathcal{F}}(-1)^{\vert S\vert +1}\prod_{i\in \cap S}(1+y_i),
		\end{equation}
		where $\cap S=\bigcap\limits_{F\in S}\textnormal{Supp}(F)$. In particular, we have
		\begin{equation*}
			\vert \Delta\vert =\sum\limits_{\emptyset\neq S\subseteq \mathcal{F}}(-1)^{\vert S\vert +1}2^{\vert \cap S\vert}.
		\end{equation*}
	\end{lemma}
	
		Let $m\in \mathbb{N}$ and let $D_i\subseteq \mathbb{F}_2^m, 1\leq i\leq 3$. Assume that $D=(1+u^2)D_1+u^2D_2+(u+u^2)D_3\subseteq \mathcal{R}^m$.\\
	   Observe that $c_{D}: \mathcal{R}^m\longrightarrow C_{D}$ defined by 
	   \begin{equation}\label{c_DMap}
	   	c_{D}(v)=\big(v\cdot d\big)_{d\in D}
	   \end{equation}
	   is an epimorphism of $\mathcal{R}$-modules. Assume that $x=(1+u^2)\alpha +u^2\beta +(u+u^2)\gamma \in \mathcal{R}^m$ and $d=(1+u^2)t_1 +u^2 t_2 +(u+u^2)t_3 \in D$, where $\alpha, \beta, \gamma \in \mathbb{F}_2^m$ and $t_i\in D_i, 1\leq i \leq 3$. Then the Lee weight of $c_{D}(x)$ is 
	   \begin{equation*}
	   	\begin{split}
	   		wt_{Lee}(c_{D}(x)) = &  ~wt_{Lee}\big(\big(\big((1+u^2)\alpha +u^2\beta +(u+u^2)\gamma \big)\cdot \big((1+u^2)t_1 +u^2t_2 +(u+u^2)t_3 \big)\big)_{t_i\in D_i}\big)\\
	   		                 = & ~wt_{Lee}\big(\big(\alpha t_1+ u(\gamma t_2 + \beta t_3) + u^2(\alpha t_1 +\beta t_2 +\gamma t_2 + \beta t_3)\big)_{t_i\in D_i}\big)\\
	   		                 = & ~wt_{H}\big(\big(\alpha t_1 +\gamma t_2 + \beta t_3\big)_{t_i\in D_i}\big) + wt_{H}\big(\big(\alpha t_1+\beta t_2\big)_{t_i\in D_i}\big) \\
	   		                 & ~+ wt_{H}\big(\big(\alpha t_1 + (\beta + \gamma)t_2 + \beta t_3\big)_{t_i\in D_i}\big).
	   	\end{split}
	   \end{equation*}   
       Now if $v \in \mathbb{F}_{2}^m$, then $wt_H(v)=0 \iff v=\textbf{0} \in \mathbb{F}_{2}^m$. Hence,
       \begin{equation}\label{keyeq1}
   	     \begin{split}
   		   wt_{Lee}(c_{D}(x))
   		     =  & ~\vert D \vert -\frac{1}{2}\sum\limits_{t_1\in D_1}\sum\limits_{t_2\in D_2}\sum\limits_{t_3\in D_3}\big(1+(-1)^{\alpha t_1 +\gamma t_2 + \beta t_3} \big)\\
   		    &+\vert D \vert -\frac{1}{2}\sum\limits_{t_1\in D_1}\sum\limits_{t_2\in D_2}\sum\limits_{t_3\in D_3}\big(1+(-1)^{\alpha t_1+\beta t_2} \big)\\
   		    &+ \vert D \vert -\frac{1}{2}\sum\limits_{t_1\in D_1}\sum\limits_{t_2\in D_2}\sum\limits_{t_3\in D_3}\big(1+(-1)^{\alpha t_1 + (\beta + \gamma)t_2 + \beta t_3} \big)\\
   		    = & ~\frac{3}{2}\vert D\vert -\frac{1}{2}\sum\limits_{t_1\in D_1}(-1)^{\alpha t_1}\big[\sum\limits_{t_2\in D_2}(-1)^{\gamma t_2}\sum\limits_{t_3\in D_3}(-1)^{\beta t_3} + \sum\limits_{t_2\in D_2}(-1)^{\beta t_2}\sum\limits_{t_3\in D_3}(1) \\
   		    &+\sum\limits_{t_2\in D_2}(-1)^{(\beta +\gamma) t_2}\sum\limits_{t_3\in D_3}(-1)^{\beta t_3} \big].
   	    \end{split}
       \end{equation}
       Note that, for $\alpha \in \mathbb{F}_2^m$ and $\emptyset \neq L \subseteq [m]$, we have
       \begin{equation}\label{keyeq2}
       	   \begin{split}
       	   	  \sum\limits_{t\in \Delta_{L}}(-1)^{\alpha t} 
       	   	  = & ~\mathcal{H}_{\Delta_{L}}\big((-1)^{\alpha_1}, (-1)^{\alpha_2},\dots , (-1)^{\alpha_m}\big)\\
       	   	  = & ~\prod_{i\in L}(1+(-1)^{\alpha_i})=\prod_{i\in L}(2-2\alpha_i)\\
       	   	  = & ~2^{\vert L \vert }\prod_{i\in L}(1-\alpha_i) = 2^{\vert L \vert}\chi(\alpha\vert L),
       	   \end{split}
       \end{equation}
      where $\chi(\cdot \vert L): \mathbb{F}_2^m \longrightarrow \mathbb{F}_2$ is the Boolean function defined by
         \begin{equation}\label{BooleanFunction}
         	\chi(\alpha\vert L)=\prod_{i\in L}(1-\alpha_i)=\begin{cases}
         		1, \text{  if } \textnormal{ Supp}(\alpha)\cap L=\emptyset;\\
         		0, \text{  if } \textnormal{ Supp}(\alpha)\cap L\neq \emptyset.
         	\end{cases}
         \end{equation}
        \begin{lemma}\label{countingLemma}
        	Suppose $L, M, N$ are subsets of $[m]$. Then
        	\begin{enumerate}
        		\item \begin{enumerate}
        		   	   \item \label{1.1}
        		   	   $\vert \{v\in \mathbb{F}_{2}^m : \chi(v \vert L) = 1\}\vert = 2^{m-\vert L\vert}$.
        			   \item \label{1.2}
        			   $\vert \{v\in \mathbb{F}_{2}^m : \chi(v \vert L) = 0\}\vert = (2^{\vert L\vert }-1)\times 2^{m-\vert L\vert}$.
        		     \end{enumerate}    		
        		\item \label{2.0}
        		$\vert \{v\in \mathbb{F}_2^m : \chi(v\vert M )=0 = \chi(v\vert N)\}\vert = (2^{\vert M\vert }-1)\times 2^{m-\vert M\vert } + (2^{\vert N\vert }-1)\times 2^{m-\vert N\vert }-(2^{\vert M\cup N\vert }-1)\times 2^{m-\vert M\cup N\vert }$.        		
        		
        		\item Fix $v\in \mathbb{F}_2^m$ with $\chi(v\vert M) = 0 $. Then 
        		\begin{enumerate}
        			\item \label{3.0}
        			$\vert \{w \in \mathbb{F}_2^m : \chi(w\vert M)=0, \chi(v+w\vert M)=0 \}\vert = (2^{\vert M\vert }-2)\times 2^{m-\vert M\vert }$,
        			
        			\item \label{3.2}
        			$\vert \{w \in \mathbb{F}_2^m : v\neq w, \chi(w\vert M)=0, \chi(v+w\vert M)=1 \}\vert = (2^{m-\vert M\vert }-1)$.
        		\end{enumerate}

        		\item \label{4.0}
        		$\vert \{(v, w)\in \mathbb{F}_2^m \times \mathbb{F}_2^m : \chi(v \vert M)=0= \chi(v\vert N) = \chi(w\vert M)= \chi(v+w\vert M) \}\vert = \big\{(2^{\vert M\vert }-1)\times 2^{m-\vert M \vert } + (2^{\vert N\vert }-1)\times 2^{m-\vert N \vert }- (2^{\vert M\cup N\vert }-1)\times 2^{m-\vert M \cup N \vert }\big\}\times \big\{(2^{\vert M\vert }-2)\times 2^{m-\vert M\vert}\big\} $.
        		
        		\item \label{5.0}
        		$\vert \{v \in \mathbb{F}_2^m : \chi(v \vert M)=0, \chi(v\vert N) = 1\}\vert = (2^{\vert M\setminus N\vert }-1)\times 2^{m-\vert M\cup N \vert }$.
        		
        		\item $\vert \{(v, w)\in \mathbb{F}_2^m \times \mathbb{F}_2^m : \chi(v \vert M)=0, \chi(v\vert N) = 1, \chi(w\vert M)=0= \chi(v+w\vert M) \}\vert = \big\{(2^{\vert M\setminus N\vert }-1)\times 2^{m-\vert M\cup N \vert }\big\} \times \big\{(2^{\vert M \vert }-2)\times 2^{m-\vert M\vert} \big\}$.
        		
        		\item $\vert \{(v, w)\in \mathbb{F}_2^m \times \mathbb{F}_2^m : v\neq w, \chi(v \vert M)=0= \chi(v\vert N) = \chi(w\vert M), \chi(v+w\vert M) = 1\}\vert = \big\{(2^{\vert M\vert }-1)\times 2^{m-\vert M \vert } + (2^{\vert N\vert }-1)\times 2^{m-\vert N \vert }- (2^{\vert M\cup N\vert }-1)\times 2^{m-\vert M \cup N \vert }\big\}\times \big\{(2^{m-\vert M\vert }-1)\big\} $.

        	\end{enumerate}
        \end{lemma}
        Here we recall a result from \cite{shi_nonchain}.
        \begin{lemma}\cite{shi_nonchain}\label{Complement_sum}
        	Let $\alpha \in \mathbb{F}_2^m$ and let $\Delta_{L}$ be the simplicial complex generated by $L \subseteq [m]$. Then
        	\begin{equation*}
        		\begin{split}
        			\sum_{t\in \Delta_{L}^c}(-1)^{\alpha t}=2^{m}\delta_{0, \alpha}-\sum_{t\in \Delta_{L}}(-1)^{\alpha t},
        		\end{split}
        	\end{equation*}
        	where $\delta$ is the Kronecker delta function and $\Delta_{L}^c=\mathbb{F}_2^m\setminus \Delta_{L}$.
        \end{lemma}
        The following results describe Lee weight distributions of $C_D$ for various choices for $D$. As proofs of all the results of the following theorem follow the same pattern so we discuss only one and omit proof for others.
		\begin{theorem}\label{main}
			Suppose that $m\in \mathbb{N}$ and $L, M, N \subseteq [m]$.
			\begin{enumerate}
				\item \label{proposition:1}
				Let $D=(1+u^2)\Delta_L+u^2\Delta_M+(u+u^2)\Delta_N\subset \mathcal{R}^m$. Then $C_D$ is a $2$-weight linear code of length $\vert D\vert = 2^{\vert L\vert + \vert M \vert +\vert N \vert}$ and size $2^{\vert L \vert + \vert M\vert +\vert M \cup N\vert}$. The Lee weight distribution of $C_D$ is displayed in Table \ref{table:1}.
					\begin{table}[]
						\centering
							\scalebox{0.7}{\begin{tabular}{  c | c  }
								\hline
								Lee weight    & Frequency \\
								\hline
								$3\times 2^{\vert L\vert+\vert M\vert+\vert N\vert-1}$ & $2^{3m}+2^{3m-\vert L\vert-\vert M\vert-\vert M\cup N\vert+1}-2^{3m-\vert L\vert-\vert M\vert}-2^{3m-\vert L\vert-\vert M\vert-\vert N\vert+1}$\\
								\hline
								$2^{ \vert L \vert + \vert M \vert + \vert N \vert}$ & $2^{3m-\vert L\vert -\vert M\vert}+2^{3m-\vert L \vert -\vert M\vert -\vert N\vert +1}-3\times 2^{3m -\vert L \vert -\vert M \vert -\vert M\cup N\vert}$\\     				
								\hline
								$0$ & $2^{3m-\vert L\vert -\vert M\vert -\vert M\cup N\vert}$\\
								\hline
							\end{tabular}}
						\caption{Lee weight distribution in Theorem \ref{main}(\ref{proposition:1})}
						\label{table:1}	
					\end{table}

				\item \label{proposition:2}
				Let $D=(1+u^2)\Delta^{c}_L+u^2\Delta_M+(u+u^2)\Delta_N\subset \mathcal{R}^m$. Then $C_D$ is a $4$-weight linear code of length $\vert D\vert = (2^m-2^{\vert L\vert})\times 2^{ \vert M \vert +\vert N \vert}$ and size $2^{m+\vert M \vert +\vert M \cup N\vert}$. The Lee weight distribution of $C_D$ is displayed in Table \ref{table:2}.
					\begin{table}[]
						\centering
							\scalebox{0.7}{\begin{tabular}{  c | c  }
								\hline
								Lee weight    & Frequency \\
								\hline
								$3\times 2^{m+\vert M\vert +\vert N\vert-1}$ & $(2^{m-\vert L\vert} -1)\times 2^{2m-\vert M\vert -\vert M\cup N\vert}$\\
								\hline
								$2^{\vert M\vert+\vert N\vert-1}(3\times 2^m-2^{\vert L \vert +1})$ & $(2^{m-\vert L \vert}-1)(2^{2m-\vert M\vert -\vert N\vert +1}+2^{2m-\vert M \vert}-3\times 2^{2m-\vert M\vert-\vert M\cup N\vert})$\\
								\hline
								$3\times (2^{m}-2^{\vert L\vert})\times 2^{\vert M\vert + \vert N\vert-1}$ & $2^{3m}-2^{3m-\vert L \vert -\vert M\vert }-2^{3m-\vert L\vert -\vert M \vert -\vert N\vert +1}+2^{3m-\vert L \vert -\vert M\vert -\vert M\cup N\vert+1}$\\
								\hline
								$(2^m-2^{\vert L\vert})\times 2^{\vert M\vert + \vert N\vert}$&$2^{2m-\vert M\vert -\vert N\vert +1}+2^{2m-\vert M\vert}-3\times 2^{2m-\vert M\vert -\vert M\cup N\vert}$\\
								\hline
								$0$&$2^{2m-\vert M\vert -\vert M\cup N\vert}$\\
								\hline
							\end{tabular}}
						\caption{Lee weight distribution in Theorem \ref{main} (\ref{proposition:2})}
						\label{table:2}		
					\end{table}
				
				\item \label{proposition:4}
				Let $D=(1+u^2)\Delta_L+u^2\Delta^{c}_M+(u+u^2)\Delta_N\subset \mathcal{R}^m$. Then $C_D$ is a $5$-weight linear code of length $\vert D\vert = (2^m-2^{\vert M \vert})\times 2^{\vert L\vert + \vert N\vert}$ and size $2^{2m+\vert L\vert }$. The Lee weight distribution of $C_D$ is displayed in Table \ref{table:4}.
				\begin{table}[]
					\centering
						\scalebox{0.7}{\begin{tabular}{  c | c  }
							\hline
							Lee weight    & Frequency \\
							\hline
							$3\times 2^{m+\vert L\vert +\vert N \vert -1}$ & $2^{3m-\vert L\vert -\vert M\vert -\vert M\cup N\vert}-2^{2m-\vert L\vert -\vert M\cup N\vert +1}-2^{2m-\vert L\vert -\vert M\vert }+2^{m-\vert L\vert +1}$\\
							\hline
							\multirow{2}{*}{$(3\times 2^{m-1}-2^{\vert M\vert})\times 2^{\vert L\vert + \vert N\vert}$} & $2^{3m-\vert L\vert -\vert M\vert -\vert N\vert +1}-2^{2m-\vert L\vert -\vert N\vert +1}+2^{3m-\vert L\vert -\vert M\vert}$\\
							 &	$-3\times 2^{3m-\vert L\vert -\vert M\vert -\vert M\cup N\vert}-2^{2m-\vert L\vert}+2^{2m-\vert L\vert -\vert M\vert}+2^{2m-\vert L\vert - \vert M\cup N\vert +1}$\\
							\hline
							$3\times (2^{m}-2^{\vert M\vert})\times 2^{\vert L\vert +\vert N\vert-1}$ & $2^{3m}-2^{3m-\vert L\vert - \vert M\vert} -2^{3m - \vert L\vert -\vert M\vert -\vert N\vert+1}+2^{3m-\vert L\vert - \vert M\vert-\vert M\cup N\vert +1}$\\
							\hline
							$2^{m+\vert L\vert +\vert N\vert}$ & $2^{2m-\vert L\vert -\vert M\cup N\vert +1} + 2^{2m-\vert L \vert -\vert M\vert}-3\times 2^{m-\vert L\vert}$\\
							\hline
							$(2^{m}-2^{\vert M\vert})\times 2^{\vert L\vert+\vert N\vert}$ & $2^{2m-\vert L\vert - \vert N\vert +1}+2^{2m-\vert L\vert}-2^{2m-\vert L\vert -\vert M\vert}-2^{2m-\vert L\vert -\vert M\cup N\vert +1}$\\
							\hline
							$0$ & $2^{m-\vert L\vert}$\\
							\hline
						\end{tabular}}
					\caption{Lee weight distribution in Theorem \ref{main} (\ref{proposition:4})}
					\label{table:4}		
				\end{table}
				
				\item \label{proposition:3}  
				Let $D=(1+u^2)\Delta_L+u^2\Delta_M+(u+u^2)\Delta^{c}_N\subset \mathcal{R}^m$. Then $C_D$ is a $4$-weight linear code of length $\vert D\vert = (2^m-2^{\vert N\vert })\times 2^{\vert L\vert +\vert M\vert}$ and size $2^{m+\vert L \vert +\vert M\vert}$. The Lee weight distribution of $C_D$ is displayed in Table \ref{table:3}.
				\begin{table}[]
					\centering
						\scalebox{0.7}{\begin{tabular}{  c | c  }
							\hline
							Lee weight    & Frequency \\
							\hline
							$2^{\vert L\vert +\vert M\vert-1}(3\times 2^m -2^{\vert N\vert +1})$ & $2^{3m-\vert L\vert -\vert M\vert -\vert N\vert+1}-2^{3m-\vert L\vert -\vert M\vert -\vert M\cup N\vert +1}$\\
							\hline
							$3\times 2^{\vert L\vert +\vert M\vert -1}(2^m-2^{\vert N\vert})$ & $2^{3m}-2^{3m-\vert L\vert -\vert M\vert}-2^{3m-\vert L\vert-\vert M\vert -\vert N\vert +1}+ 2^{3m-\vert L\vert -\vert M\vert -\vert M\cup N\vert +1}$\\
							\hline
							$2^{m+\vert L\vert +\vert M\vert}$ & $2^{3m-\vert L\vert -\vert M\vert -\vert M\cup N\vert} - 2^{2m-\vert L\vert -\vert M\vert}$\\
							\hline
							$2^{\vert L\vert +\vert M\vert}\times (2^{m}-2^{\vert N\vert })$ & $2^{3m-\vert L\vert -\vert M\vert}-2^{3m-\vert L\vert -\vert M\vert -\vert M\cup N\vert}$\\
							\hline
							$0$ & $2^{2m-\vert L\vert -\vert M\vert}$\\
							\hline
						\end{tabular}}
					\caption{Lee weight distribution in Theorem \ref{main} (\ref{proposition:3})}
					\label{table:3}		
				\end{table}

				\item \label{proposition:5}
				Let $D=(1+u^2)\Delta^{c}_L+u^2\Delta^{c}_M+(u+u^2)\Delta_N\subset \mathcal{R}^m$. Then $C_D$ is a $10$-weight linear code of length $\vert D\vert = (2^m-2^{\vert L\vert })(2^m-2^{\vert M\vert})\times 2^{\vert N\vert}$ and size $2^{3m}$. The Lee weight distribution of $C_D$ is displayed in Table \ref{table:5}.
				\begin{table}[]
					\centering
						\scalebox{0.7}{\begin{tabular}{  c | c  }
							\hline
							Lee weight    & Frequency \\
							\hline
							$3(2^m-2^{\vert  L\vert })\times 2^{m+ \vert  N\vert -1}$ & $2^{2m-\vert  M\vert -\vert  M\cup N\vert }-2^{m-\vert  M\vert }-2^{m-\vert  M\cup N\vert +1} +2$\\
							\hline
							$3(2^m-2^{\vert  M\vert })\times 2^{m+\vert  N\vert -1}$ & $2^{m-\vert  L\vert }-1$\\
							\hline
							\multirow{2}{*}{$(2^m-2^{\vert  L\vert })(3\times 2^{m-1}-2^{\vert  M\vert })\times 2^{\vert  N\vert }$} & $2^{m-\vert  M\vert }-2^{m-\vert  N\vert +1}+2^{2m-\vert  M\vert -\vert  N\vert +1}$\\ & $-3\times 2^{2m-\vert  M\vert -\vert  M\cup N\vert }+2^{2m-\vert  M\vert }-2^m+2^{m-\vert  M\cup N\vert +1}$\\
							\hline
							\multirow{2}{*}{$(3\times 2^{m-1}-2^{\vert  L\vert })(2^{m}-2^{\vert  M\vert })\times 2^{\vert  N\vert }$} & $2^{2m-\vert  L\vert -\vert  N\vert +1}- 2^{m-\vert  N\vert +1}-2^{2m-\vert  L\vert -\vert  M\cup N\vert +1}$\\ &$+2^{m-\vert  M\cup N\vert +1}+2^{2m-\vert  L\vert }-2^{m}-2^{2m-\vert  L\vert -\vert  M\vert }+2^{m-\vert  M\vert }$\\
							\hline
							\multirow{2}{*}{$3\times 2^{m+\vert  N\vert -1}(2^m-2^{\vert  M\vert })-2^{m+\vert L \vert +\vert N \vert }$} & $2^{2m-\vert  L\vert -\vert M\cup N \vert +1}-2^{m-\vert  M\cup N\vert +1}$\\ & $-3\times 2^{m-\vert  L\vert }-2^{m-\vert  M\vert }+2^{2m-\vert  L\vert -\vert  M\vert }+3$\\
							\hline
							$3\times (2^m-2^{\vert  L\vert })(2^m-2^{\vert  M\vert })\times 2^{\vert  N\vert -1}$ & $2^{3m}-2^{3m-\vert  L\vert -\vert  M\vert }-2^{3m-\vert  L\vert -\vert  M\vert -\vert  N\vert +1}+2^{3m-\vert  L\vert -\vert  M\vert -\vert  M\cup N\vert +1}$\\
							\hline
							\multirow{4}{*}{$3\times 2^{m+\vert  N\vert -1}(2^{m}-2^{\vert  L\vert }-2^{\vert  M\vert })+2^{\vert L \vert +\vert  M\vert +\vert  N\vert }$} & $2^{3m-\vert  L\vert -\vert  M\vert -\vert  N\vert +1}-2^{2m-\vert  L\vert -\vert  N\vert +1}-2^{2m-\vert  M\vert -\vert  N\vert +1}+2^{m-\vert  N\vert +1}$\\&$+2^{2m-\vert  L\vert -\vert  M\cup N\vert +1}-2^{m-\vert  M\cup N\vert +1}+2^{3m-\vert  L\vert -\vert M \vert }-2^{2m-\vert  M\vert }$\\&$-2^{2m-\vert  L\vert }+2^m-3\times 2^{3m-\vert  L\vert -\vert  M\vert -\vert  M\cup N\vert }+2^{2m-\vert  L\vert -\vert  M\vert }$\\&$+3\times 2^{2m-\vert  M\vert -\vert  M\cup N\vert }-2^{m-\vert  M\vert }$\\
							\hline
							\multirow{2}{*}{$3(2^m-2^{\vert  L\vert }-2^{\vert  M\vert })\times 2^{m+\vert  N\vert -1}$} & $2^{3m-\vert  L\vert -\vert  M\vert -\vert  M\cup N\vert }-2^{2m-\vert L \vert -\vert M \vert }-2^{2m-\vert  M\vert -\vert M\cup N \vert } +2^{m-\vert  M\vert }$\\ &$-2^{2m-\vert L \vert -\vert  M\cup N\vert +1}+2^{m-\vert  L\vert +1}+2^{m-\vert M\cup N \vert +1 }-2$\\
							\hline
							$(2^m-2^{\vert L \vert})\times 2^{m+\vert N \vert}$&$2^{m-\vert M\cup N \vert +1}+2^{m-\vert M \vert}-3$\\
							\hline
							$(2^{m}-2^{\vert L \vert})(2^m-2^{\vert M \vert})\times 2^{\vert N \vert}$&$2^{m}+2^{m-\vert N \vert +1}-2^{m-\vert  M\vert}-2^{m-\vert M\cup N \vert +1}$\\
							\hline
							$0$&$1$\\
							\hline
						\end{tabular}}
					\caption{Lee weight distribution in Theorem \ref{main} (\ref{proposition:5})}
					\label{table:5}		
				\end{table}

			\item \label{proposition:6}
			Let $D=(1+u^2)\Delta^{c}_L+u^2\Delta_M+(u+u^2)\Delta^{c}_N\subset \mathcal{R}^m$. Then $C_D$ is a $8$-weight linear code of length $\vert D\vert = (2^m-2^{\vert L\vert })(2^m-2^{\vert N\vert })\times 2^{\vert M\vert }$ and size $2^{2m+\vert M\vert}$. The Lee weight distribution of $C_D$ is displayed in Table \ref{table:6}.
			\begin{table}[]
				\centering
					\scalebox{0.6}{\begin{tabular}{  c | c  }
						\hline
						Lee weight    & Frequency \\
						\hline
						$(2^m-2^{\vert L \vert })(3\times 2^m-2^{\vert N \vert +1})\times 2^{\vert M \vert -1}$ & $2^{2m-\vert  M\vert -\vert  N\vert +1}-2^{2m-\vert  M\vert -\vert  M\cup N\vert +1}$\\
						\hline
						$(2^m-2^{\vert L \vert })\times 2^{m+\vert M \vert }$ & $2^{2m-\vert  M\vert -\vert M\cup N \vert }-2^{m-\vert M \vert }$\\
						\hline
						$3\times (2^m-2^{\vert N \vert })\times 2^{m+\vert M \vert -1}$ & $2^{2m-\vert  L\vert -\vert  M\vert }-2^{m-\vert  M\vert }$\\
						\hline
						$(3\times 2^m-2^{\vert L \vert +1})(2^m-2^{\vert N \vert })\times 2^{\vert M \vert -1}$ & $2^{3m-\vert  L\vert -\vert  M\vert }-2^{2m-\vert M \vert }-2^{3m-\vert L \vert -\vert M \vert -\vert M\cup N \vert }+2^{2m-\vert M \vert -\vert  M\cup N\vert }$\\
						\hline
						$(3\times 2^m-2^{\vert L \vert +1})(2^m-2^{\vert N \vert })\times 2^{\vert M \vert -1}-2^{\vert L \vert +\vert M \vert +\vert N \vert }$ & $2^{3m-\vert L \vert -\vert M \vert -\vert M\cup N \vert }-2^{2m-\vert M \vert -\vert M\cup N \vert }-2^{2m-\vert  L\vert -\vert M \vert }+2^{m-\vert M \vert }$\\
						\hline
						$3(2^m-2^{\vert L \vert })(2^m-2^{\vert N \vert })\times 2^{\vert M \vert -1}$ & $2^{3m}-2^{3m-\vert L \vert -\vert  M\vert }-2^{3m-\vert L \vert -\vert M \vert -\vert N \vert +1}+2^{3m-\vert  L\vert -\vert  M\vert -\vert  M\cup N\vert +1}$\\
						\hline
						\multirow{2}{*}{$3(2^m-2^{\vert L \vert })(2^m-2^{\vert N \vert })\times 2^{\vert M \vert -1}-2^{\vert L \vert +\vert M \vert +\vert  N\vert -1}$} & $2^{3m-\vert  L\vert -\vert M \vert -\vert N \vert +1}-2^{2m-\vert M \vert -\vert N \vert +1}-2^{3m-\vert  L\vert -\vert M \vert -\vert  M\cup N\vert +1}$\\ & $+2^{2m-\vert  M\vert -\vert  M\cup N\vert +1}$\\
						\hline
						$(2^{m}-2^{\vert L \vert })(2^m-2^{\vert N \vert })\times 2^{\vert M \vert }$&$2^{2m-\vert  M\vert }-2^{2m-\vert M \vert -\vert M\cup N \vert }$\\
						\hline
						$0$&$2^{m-\vert  M\vert }$\\
						\hline
					\end{tabular}}
				\caption{Lee weight distribution in Theorem \ref{main} (\ref{proposition:6})}
				\label{table:6}		
			\end{table}
			
			\item \label{proposition:7}
			Let $D=(1+u^2)\Delta_L+u^2\Delta^{c}_M+(u+u^2)\Delta^{c}_N\subset \mathcal{R}^m$. Then $C_D$ is a $8$-weight linear code of length $\vert D\vert = (2^m-2^{\vert M\vert})(2^m-2^{\vert N\vert})\times 2^{\vert L\vert}$ and size $2^{2m+\vert L\vert}$. The Lee weight distribution of $C_D$ is displayed in Table \ref{table:7}.
			\begin{table}[]
				\centering
					\scalebox{0.6}{\begin{tabular}{  c | c  }
						\hline
						Lee weight    & Frequency \\
						\hline
						$(2^m-2^{\vert  M\vert })(3\times 2^m-2^{\vert  N\vert +1})\times 2^{\vert  L\vert -1}$ &  \multirow{2}{*}{$2^{2m-\vert L \vert -\vert M\cup N \vert +1}-2^{m-\vert L \vert +1}$}\\					
						 $+ (2^m-2^{\vert  N\vert })\times 2^{\vert L \vert +\vert M \vert -1}-2^{\vert L \vert +\vert  M\vert +\vert N \vert -1}$ &\\
						\hline
						$(2^{m}-2^{\vert N \vert })\times 2^{m+\vert L \vert }$ & $2^{2m-\vert L \vert -\vert M \vert }-2^{m-\vert L \vert }$\\
						\hline
						$(2^m-2^{\vert M \vert })(3\times 2^m-2^{\vert N \vert +1})\times 2^{\vert L \vert -1}$ & $2^{2m-\vert L \vert -\vert  N\vert +1}-2^{2m-\vert L \vert -\vert M\cup N \vert +1}$\\
						\hline
						$(3\times 2^m-2^{\vert M \vert +1})(2^m-2^{\vert N \vert })\times 2^{\vert L \vert -1}$ & $2^{3m-\vert L \vert -\vert  M\vert }-2^{2m-\vert L \vert }+2^{2m-\vert L \vert -\vert M \vert }-2^{3m-\vert  L\vert -\vert  M\vert -\vert M\cup N \vert }$\\
						\hline
						$(3\times 2^m-2^{\vert M \vert +1})(2^m-2^{\vert N \vert })\times 2^{\vert L \vert -1}-2^{\vert L \vert +\vert M \vert +\vert N \vert }$ & $2^{3m-\vert L \vert -\vert M \vert -\vert M\cup N \vert }-2^{2m-\vert L \vert -\vert M \vert }-2^{2m-\vert L \vert -\vert M\cup N \vert +1}+2^{m-\vert L \vert +1}$\\
						\hline
						$3(2^m-2^{\vert M \vert })(2^m-2^{\vert N \vert })\times 2^{\vert L \vert -1}$ & $2^{3m}-2^{3m-\vert L \vert -\vert  M\vert }-2^{3m-\vert L \vert -\vert M \vert -\vert N \vert +1}+2^{3m-\vert L \vert -\vert M \vert -\vert M\cup N \vert +1}$\\
						\hline
						\multirow{2}{*}{$3(2^m-2^{\vert M \vert })(2^m-2^{\vert N \vert })\times 2^{\vert L \vert -1}-2^{\vert L \vert +\vert M \vert +\vert N \vert -1}$ } & $2^{3m-\vert L \vert -\vert M \vert -\vert N \vert +1}-2^{3m-\vert L \vert -\vert M \vert -\vert M\cup N \vert +1}-2^{2m-\vert  L\vert -\vert N \vert +1}$\\ & $+2^{2m-\vert L \vert -\vert M\cup N \vert +1}$\\
						\hline
						$(2^m-2^{\vert M \vert })(2^m-2^{\vert N \vert })\times 2^{\vert L \vert }$&$2^{2m-\vert L \vert }-2^{2m-\vert L \vert -\vert M \vert }$\\
						\hline
						$0$&$2^{m-\vert L \vert }$\\
						\hline
					\end{tabular}}
				\caption{Lee weight distribution in Theorem \ref{main} (\ref{proposition:7})}
				\label{table:7}		
			\end{table}
			
			\item \label{proposition:8}
			Let $D=(1+u^2)\Delta^{c}_L+u^2\Delta^{c}_M+(u+u^2)\Delta^{c}_N\subset \mathcal{R}^m$. Then $C_D$ is a $16$-weight linear code of length $\vert D\vert = (2^m-2^{\vert L\vert })(2^m-2^{\vert M\vert})(2^m-2^{\vert N\vert })$ and size $2^{3m}$. The Lee weight distribution of $C_D$ is displayed in Table \ref{table:8}.
			\begin{table}[]
				\centering
					\scalebox{0.6}{\begin{tabular}{  c | c  }
						\hline
						Lee weight    & Frequency \\
						\hline
						$(2^m-2^{\vert L \vert })(2^m-2^{\vert M \vert })(3\times 2^{m-1}-2^{\vert N \vert })$ & \multirow{2}{*}{$2^{m-\vert M\cup N \vert +1}-2$}\\
						$+(2^{m}-2^{\vert L \vert })(2^m-2^{\vert N \vert +1})\times 2^{\vert M \vert -1}$ & \\
						\hline
						$3\times (2^{m}-2^{\vert M \vert })(2^m-2^{\vert N \vert })\times 2^{m-1}$ & $2^{m-\vert L \vert }-1$\\
						\hline
						$(2^{m}-2^{\vert L \vert })(2^{m}-2^{\vert M \vert })(3\times 2^{m-1}-2^{\vert N \vert })$ & $2^{m-\vert N \vert +1}-2^{m-\vert  M\cup N\vert +1}$\\
						\hline
						$(2^m-2^{\vert L \vert })(3\times 2^{m-1}-2^{\vert M \vert })(2^m-2^{\vert N \vert })$ & $2^{m-\vert  M\vert }+2^{2m-\vert  M\vert }-2^m-2^{2m-\vert  M\vert -\vert M\cup N \vert }$\\
						\hline
						\multirow{3}{*}{$3(2^m-2^{\vert L \vert })(2^m-2^{\vert M \vert })(2^{m-1}-2^{\vert N \vert -1})+2^{\vert L \vert +\vert M \vert +\vert N \vert -1}$} & $2^{3m-\vert L \vert -\vert M \vert -\vert N \vert +1}-2^{2m-\vert L \vert -\vert N \vert +1}-2^{2m-\vert  M\vert -\vert  N\vert +1}$\\
						 & $+2^{m-\vert N \vert +1}-2^{3m-\vert  L\vert -\vert  M\vert -\vert  M\cup N\vert +1}+2^{2m-\vert L \vert -\vert M\cup N \vert +1}$\\
						  & $+2^{2m-\vert M \vert -\vert M\cup N \vert +1}-2^{m-\vert M\cup N \vert +1}$\\
						\hline
						$(3\times 2^{m-1}-2^{\vert L \vert })(2^m-2^{\vert M \vert })(2^m-2^{\vert N \vert })$ & $2^{2m-\vert L \vert }-2^m-2^{2m-\vert L \vert -\vert M \vert }+2^{m-\vert M \vert }$\\
						\hline
						$(2^m-2^{\vert L \vert })(3\times 2^{m-1}-2^{\vert M \vert })(2^m-2^{\vert N\vert }) -(2^m - 2^{\vert L \vert})2^{\vert M \vert + \vert N \vert}$ & $2^{2m-\vert  M\vert -\vert M\cup N \vert }-2^{m-\vert  M\vert }-2^{m-\vert M\cup N \vert +1}+2$\\
						\hline
						$(3\times 2^{m-1}-2^{\vert L \vert })(2^m-2^{\vert M \vert })(2^m-2^{\vert  N\vert })-(2^{m}-2^{\vert N \vert })\times 2^{\vert L \vert +\vert M \vert }$ & $2^{2m-\vert  L\vert -\vert M \vert }-2^{m-\vert M \vert }-2^{m-\vert L \vert }+1$\\
						\hline
						$3(2^m-2^{\vert L \vert })(2^m-2^{\vert M \vert })(2^{m-1}-2^{\vert N \vert -1})$ & $2^{3m-\vert L \vert -\vert  M\vert -\vert M\cup N \vert }-2^{2m-\vert M \vert -\vert M\cup N \vert }-2^{2m-\vert  L\vert -\vert  M\vert }$\\
						$-(2^m-2^{\vert N \vert })2^{\vert  L\vert +\vert  M\vert -1}+2^{\vert  L\vert +\vert  M\vert +\vert  N\vert }$ & $+2^{m-\vert M \vert }-2^{2m-\vert  L\vert -\vert M\cup N \vert +1}+2^{m-\vert M\cup N \vert +1}+2^{m-\vert L \vert +1}-2$\\
						\hline
						$3(2^m-2^{\vert L \vert })(2^m-2^{\vert M \vert })(2^{m-1}-2^{\vert N \vert -1})-(2^m-2^{\vert M \vert })2^{\vert  L\vert +\vert N \vert -1}$ & \multirow{2}{*}{$2^{2m-\vert L \vert -\vert M\cup N \vert +1}-2^{m-\vert M\cup N \vert +1}-2^{m-\vert L \vert +1}+2$}\\
						$-(2^m-2^{\vert N \vert })2^{\vert  L\vert +\vert M \vert -1}+2^{\vert L \vert +\vert  M\vert +\vert N \vert -1}$ & \\
						\hline
						$3\times (2^m-2^{\vert L \vert })(2^m-2^{\vert M \vert })(2^{m-1}-2^{\vert  N\vert -1})-(2^m-2^{\vert L \vert })\times 2^{\vert  M\vert +\vert N \vert -1}$ & $2^{2m-\vert M \vert -\vert N \vert +1}-2^{2m-\vert M \vert -\vert M\cup N \vert +1}-2^{m-\vert N \vert +1}+2^{m-\vert M\cup N \vert +1}$\\
						\hline
						$3\times (2^m-2^{\vert L \vert })(2^m-2^{\vert  M\vert })(2^{m-1}-2^{\vert N \vert -1})-(2^m-2^{\vert  M\vert })\times 2^{\vert L \vert +\vert N \vert -1}$ & $2^{2m-\vert  L\vert -\vert N \vert +1}-2^{m-\vert  N\vert +1}-2^{2m-\vert  L\vert -\vert M\cup N \vert +1}+2^{m-\vert M\cup N \vert +1}$\\
						\hline
						\multirow{2}{*}{$3\times (2^m-2^{\vert L \vert })(2^m-2^{\vert M \vert })(2^{m-1}-2^{\vert N \vert -1})-(2^m-2^{\vert N \vert })\times 2^{\vert L \vert +\vert  M\vert -1}$} & $2^m-2^{m-\vert M \vert }+2^{3m-\vert L \vert -\vert M \vert }-2^{2m-\vert M \vert }-2^{2m-\vert L \vert }$\\ & $-2^{3m-\vert L \vert -\vert M \vert -\vert M\cup N \vert }+2^{2m-\vert M \vert -\vert  M\cup N\vert }+2^{2m-\vert L \vert -\vert M \vert }$\\
						\hline
						$(2^m-2^{\vert L \vert })(2^m-2^{\vert N \vert })\times 2^m$ & $2^{m-\vert M \vert }-1$\\
						\hline
						$3\times (2^m-2^{\vert L \vert })(2^m-2^{\vert M \vert })(2^{m-1}-2^{\vert N \vert -1})$ & $2^{3m}-2^{3m-\vert L \vert -\vert M \vert }-2^{3m-\vert  L\vert -\vert M \vert -\vert N \vert +1}+2^{3m-\vert  L\vert -\vert M \vert -\vert M\cup N \vert +1}$\\
						\hline
						$(2^{m}-2^{\vert L \vert })(2^m-2^{\vert M \vert })(2^m-2^{\vert N \vert })$ & $2^m-2^{m-\vert M \vert }$\\
						\hline
						$0$ & $1$\\
						\hline
					\end{tabular}}
				\caption{Lee weight distribution in Theorem \ref{main} (\ref{proposition:8})}
				\label{table:8}		
			\end{table}
				
			\end{enumerate}
		\end{theorem}		
		 
         \proof Now we prove part (\ref{proposition:8}).\\ 
         Observe that $C_{D}$ has length $\vert D\vert = (2^m-2^{\vert L\vert})(2^m-2^{\vert M\vert})(2^m-2^{\vert N\vert})$. Let $x = (1+u^2)\alpha +u^2\beta + (u+u^2)\gamma \in \mathcal{R}^m$. By Equation \eqref{keyeq1} and \eqref{keyeq2}, we have
         \begin{equation}
         	\begin{split}
         		wt_{Lee}(c_{D}(x)) = & ~\frac{3}{2}\vert D\vert -\frac{1}{2}(2^m\delta_{0, \alpha}-2^{\vert L\vert }\chi(\alpha \vert L))\big[(2^m \delta_{0, \gamma} - 2^{\vert M\vert }\chi(\gamma \vert M))(2^m\delta_{0, \beta}-2^{\vert N\vert }\chi(\beta \vert N))\\
         		& + (2^m\delta_{0, \beta} - 2^{\vert M\vert }\chi(\beta \vert M))(2^m-2^{\vert N\vert }) \\
         		& + (2^m\delta_{0, \beta+\gamma}-2^{\vert M\vert }\chi(\beta +\gamma \vert M))(2^m \delta_{0, \beta}- 2^{\vert N\vert }\chi(\beta \vert N))\big].
         	\end{split}
         \end{equation}
        Now we look at the following cases.
        \begin{enumerate}
        	\item If $\alpha =0, \beta =0, \gamma =0\implies \beta +\gamma =0$ then\\
        	\begin{equation*}
        		\begin{split}
        			wt_{Lee}(c_D(x)) =  & ~0.
        		\end{split}
        	\end{equation*}
            Here in this case, $\#\alpha =1, \#\beta =1, \#\gamma =1$. Therefore, $\# x = 1$.
            \item If $\alpha =0, \beta \neq 0, \gamma =0 \implies \beta +\gamma \neq 0$ then\\
            \begin{equation*}
            	\begin{split}
                  wt_{Lee}(c_D(x)) = & ~\frac{3}{2}\vert D\vert -\frac{1}{2}(2^m-2^{\vert L\vert })\big[-(2^m-2^{\vert M\vert })2^{\vert N\vert }\chi(\beta \vert N) - 2^{\vert M\vert }(2^m-2^{\vert N\vert })\chi(\beta \vert M)\\
                  & + 2^{\vert M\vert +\vert N\vert}\chi(\beta \vert M)\chi(\beta \vert N)\big]
            	\end{split}
            \end{equation*}
            \begin{itemize}
            	\item If $\chi(\beta \vert M) = 0, \chi(\beta \vert N) = 0$ then $wt_{Lee}(c_D(x))=\frac{3}{2}\vert D\vert$.\\
            	In this case, by using Lemma \ref{countingLemma},  we get\\
            	 $\#\alpha =1$, $\#\gamma = 1$,\\
            	$\#\beta = (2^{\vert M\vert }-1)\times 2^{m-\vert M \vert } + (2^{\vert N\vert }-1)\times 2^{m-\vert N \vert }- (2^{\vert M\cup N\vert }-1)\times 2^{m-\vert M \cup N \vert }$.\\
            	 Therefore, $\#x = (2^{\vert M\vert }-1)\times 2^{m-\vert M \vert } + (2^{\vert N\vert }-1)\times 2^{m-\vert N \vert }- (2^{\vert M\cup N\vert }-1)\times 2^{m-\vert M \cup N \vert }$.
            	
            	\item If $\chi(\beta \vert M) = 0, \chi(\beta \vert N) = 1$ then 
            	$wt_{Lee}(c_D(x)) = \frac{3}{2}\vert D\vert + \frac{1}{2}(2^m-2^{\vert L\vert })(2^m-2^{\vert M\vert })\times2^{\vert N\vert}$.
            	In this case, by using Lemma \ref{countingLemma},  we get \\
            	$\#\alpha =1$,
            	$\#\beta = (2^{\vert M\setminus N\vert}-1)\times 2^{m-\vert M \cup N\vert}$,
            	$\#\gamma = 1$.\\
            	Therefore, $\#x= (2^{\vert M\setminus N\vert}-1)\times 2^{m-\vert M \cup N\vert}$.

            	\item If $\chi(\beta \vert M) = 1, \chi(\beta \vert N) = 0$ then $wt_{Lee}(c_D(x))=\frac{3}{2}\vert D\vert +\frac{1}{2}(2^m-2^{\vert L \vert })(2^m-2^{\vert N\vert })\times 2^{\vert M\vert}$.
            	In this case, by using Lemma \ref{countingLemma},  we get \\
            	$\#\alpha =1$,
            	$\#\beta = (2^{\vert N\setminus M\vert}-1)\times 2^{m-\vert M \cup N\vert}$,
            	$\#\gamma = 1$.\\
            	Therefore, $\#x= (2^{\vert N\setminus M\vert}-1)\times 2^{m-\vert M \cup N\vert}$.
            	\item If $\chi(\beta \vert M) = 1, \chi(\beta \vert N) = 1$ then $wt_{Lee}(c_D(x)) = \frac{3}{2}\vert D\vert +\frac{1}{2}(2^m-2^{\vert L\vert})(2^m-2^{\vert M\vert})\times 2^{\vert N\vert}+\frac{1}{2}(2^m-2^{\vert L\vert})(2^m-2^{\vert N\vert})\times 2^{\vert M\vert} -\frac{1}{2}(2^m-2^{\vert L\vert})\times 2^{\vert M\vert + \vert N\vert}$. In this case, by using Lemma \ref{countingLemma},  we get \\
            	$\#\alpha =1$,
            	$\#\beta = 2^{m-\vert M \cup N\vert}-1$,
            	$\#\gamma = 1$.\\
            	Therefore, $\#x= 2^{m-\vert M \cup N\vert}-1$.
            \end{itemize}
         \end{enumerate}
     Similarly, we can explore the cases: $\alpha = 0, \beta = 0, \gamma \neq 0$; $\alpha = 0, \beta \neq 0, \gamma \neq 0, \beta =\gamma $; $\alpha = 0, \beta \neq 0, \gamma \neq 0, \beta \neq \gamma $; $\alpha \neq 0, \beta =0, \gamma =0$; $\alpha \neq 0, \beta \neq 0, \gamma =0 $; $\alpha \neq 0, \beta = 0, \gamma \neq 0 $; $\alpha \neq 0, \beta \neq 0, \gamma \neq 0, \beta =\gamma $ and $\alpha \neq 0, \beta \neq 0, \gamma \neq 0, \beta \neq \gamma $.\\
         Based on the above calculations, we\label{key} obtain Table \ref{table:8}.\\
         Consider the map $c_{D}:\mathcal{R}^m\longrightarrow C_{D}$ defined by $c_{D}(v)=\big(v\cdot d\big)_{d\in D}$ which is a surjective linear transformation.
         By using Table \ref{table:8}, $\vert \ker(c_D)\vert = \vert \{v\in \mathcal{R}^m : v\cdot d = 0 ~\forall ~d \in D\}\vert =1$ so that $c_D$ is an isomorphism and hence $\vert C_{D}\vert =2^{3m}$.\qed         
        
  
        \subsection{Using simplicial complexes having two maximal elements}
        In this subsection, we discuss codes constructed using simplicial complexes with two maximal elements. First we recall a result from [\cite{Wu_Li}, Lemma 4.1].
        \begin{lemma}\cite{Wu_Li}\label{U_iLemma}
        	For two subsets $L$, $M$ of $[m]$, we set
        	\begin{enumerate}
        		\item $\mathfrak{U}_1 =\{v \in \mathbb{F}_2^m : v \cap (L \cup M) = \emptyset \}$,
        		\item $\mathfrak{U}_2 =\{v \in \mathbb{F}_2^m : v\cap L = \emptyset, v \cap (M \setminus L) \neq \emptyset \}$,
        		\item $\mathfrak{U}_3 =\{v \in \mathbb{F}_2^m : v\cap M = \emptyset, v \cap (L \setminus M) \neq \emptyset \}$,
        		\item $\mathfrak{U}_4 =\{v \in \mathbb{F}_2^m :  v \cap (L \setminus M) \neq \emptyset ,  v \cap (M \setminus L) \neq \emptyset,  v \cap (L \cap M) = \emptyset\}$,
        		\item $\mathfrak{U}_5 =\{v \in \mathbb{F}_2^m :  v \cap (L \cap M) \neq \emptyset\}$.
        	\end{enumerate}
        	Then we have,
        	\begin{enumerate}
        		\item $\vert \mathfrak{U}_1 \vert = 2^{m-\vert L \cup M\vert}$ ,
        		\item $\vert \mathfrak{U}_2 \vert = (2^{\vert M \setminus L \vert} -1)2^{m-\vert L \cup M\vert}$,
        		\item $\vert \mathfrak{U}_3 \vert = (2^{\vert L \setminus M \vert} -1)2^{m-\vert L \cup M\vert}$,
        		\item $\vert \mathfrak{U}_4 \vert = (2^{\vert L \setminus M \vert} -1)(2^{\vert M \setminus L \vert} -1)2^{m-\vert L \cup M\vert}$,
        		\item $\vert \mathfrak{U}_5 \vert = (2^{\vert L \cap M \vert} -1)2^{m-\vert L \cap M\vert}$.
        	\end{enumerate}
        \end{lemma}
        
        Let $\Delta$ be the simplicial complex having two maximal elements $L$ and $M$, where $L, M\subseteq [m]$. Now we have the following result regarding linear codes over $\mathcal{R}$ constructed using $\Delta$.
        \begin{theorem}\label{mainTwo}
        	Suppose that $m\in \mathbb{N}$ and $L, M, N \subseteq [m]$.
        	\begin{enumerate}
        		\item \label{proposition:1Two}
        		Let $D=(1+u^2)\Delta+u^2\Delta+(u+u^2)\Delta \subseteq \mathcal{R}^m$. Then $C_D$ is a linear code of length $\vert D\vert = (2^{\vert L\vert} + 2^{\vert M \vert } -2^{\vert L \cap M \vert })^3$ and size $2^{3\vert L  \cup  M\vert }$.
        		
        		\item \label{proposition:2Two}
        		Let $D=(1+u^2)\Delta^{c}+u^2\Delta+(u+u^2)\Delta \subseteq \mathcal{R}^m$. Then $C_D$ is a linear code of length $\vert D\vert = (2^m - 2^{\vert L\vert} - 2^{\vert M \vert } +2^{\vert L \cap M \vert })(2^{\vert L\vert} + 2^{\vert M \vert } -2^{\vert L \cap M \vert })^2$ and size $2^{m + 2\vert L\cup M \vert }$. 
        		
        		\item \label{proposition:3Two}
        		Let $D=(1+u^2)\Delta + u^2\Delta^{c} + (u+u^2)\Delta \subseteq \mathcal{R}^m$. Then $C_D$ is a linear code of length $\vert D\vert = (2^m - 2^{\vert L\vert} - 2^{\vert M \vert } +2^{\vert L \cap M \vert })(2^{\vert L\vert} + 2^{\vert M \vert } -2^{\vert L \cap M \vert })^2$ and size $2^{2m+\vert L \cup M\vert }$. 
        		
        		\item \label{proposition:4Two}
        		Let $D=(1+u^2)\Delta + u^2\Delta +(u+u^2)\Delta^{c} \subseteq \mathcal{R}^m$. Then $C_D$ is a linear code of length $\vert D\vert = (2^m - 2^{\vert L\vert} - 2^{\vert M \vert } +2^{\vert L \cap M \vert })(2^{\vert L\vert} + 2^{\vert M \vert } -2^{\vert L \cap M \vert })^2$ and size $2^{m + 2\vert L \cup M \vert }$.
        		
        		\item \label{proposition:5Two}
        		Let $D=(1+u^2)\Delta^{c} +u^2\Delta^{c} +(u+u^2)\Delta \subseteq \mathcal{R}^m$. Then $C_D$ is a linear code of length $\vert D\vert = (2^m - 2^{\vert L\vert} - 2^{\vert M \vert } +2^{\vert L \cap M \vert })^2 (2^{\vert L\vert} + 2^{\vert M \vert } -2^{\vert L \cap M \vert })$ and size $2^{3m}$.
        		
        		\item \label{proposition:6and7Two}
        		Let $D=(1+u^2)\Delta^{c} +u^2\Delta +(u+u^2)\Delta^{c}$ or $(1+u^2)\Delta +u^2\Delta^{c} +(u+u^2)\Delta^{c} \subseteq \mathcal{R}^m$. Then $C_D$ is a linear code of length $\vert D\vert = (2^m - 2^{\vert L\vert} - 2^{\vert M \vert } +2^{\vert L \cap M \vert })^2 (2^{\vert L\vert} + 2^{\vert M \vert } -2^{\vert L \cap M \vert })$ and size $2^{2m+\vert L \cup M\vert}$.
        	
        		\item \label{proposition:8Two}
        		Let $D=(1+u^2)\Delta^{c} +u^2\Delta^{c} +(u+u^2)\Delta^{c} \subseteq \mathcal{R}^m$. Then $C_D$ is a linear code of length $\vert D\vert = (2^m - 2^{\vert L\vert} - 2^{\vert M \vert } +2^{\vert L \cap M \vert })^3$ and size $2^{3m}$.
        		
        	\end{enumerate}
        \end{theorem}		
        
        \proof Now we prove part (\ref{proposition:1Two}).\\
        For $\alpha \in \mathbb{F}_2^m$ and $P\subseteq \mathbb{F}_2^m$, we define
        \begin{equation}\label{DefinePsiFunction}
        	\Psi_{\alpha}(P) = \sum_{y\in P} (-1)^{\alpha \cdot y}.
        \end{equation}
        By Lemma \ref{generatinglemma}, we have
        \begin{equation}\label{keyeqTwo}
        	\begin{split}
        		\Psi_{\alpha}(\Delta) & = \sum_{y \in \Delta}(-1)^{\alpha \cdot y} = \mathcal{H}((-1)^{\alpha_1}, (-1)^{\alpha_2}, \dots , (-1)^{\alpha_m})\\
        		& = \prod_{i\in L}(1 + (-1)^{\alpha_i}) + \prod_{i\in M}(1 + (-1)^{\alpha_i}) - \prod_{i\in L\cap M}(1 + (-1)^{\alpha_i})\\
        		& = 2^{\vert L \vert }\chi(\alpha \vert L) + 2^{\vert M \vert}\chi(\alpha \vert M) - 2^{\vert L \cap M \vert}\chi(\alpha \vert L \cap M ),
        	\end{split}
        \end{equation}
        where $\chi$ is the Boolean function defined by Equation \eqref{BooleanFunction}.\\
        In particular, for $y=0 \in \mathbb{F}_2^m$, we have $\vert \Delta \vert = 2^{\vert L \vert} +2^{\vert M \vert }-2^{\vert L \cap M \vert}$. So the length of the code $C_{D}$ is $\vert D \vert = (2^{\vert L\vert} + 2^{\vert M \vert } -2^{\vert L \cap M \vert })^3$.
        Let $x = (1+u^2)\alpha +u^2\beta + (u+u^2)\gamma \in \mathcal{R}^m$. By Equation \eqref{keyeq1} and \eqref{keyeqTwo}, we have
        \begin{equation}\label{Eq3.12}
        	\begin{split}
        		wt_{Lee}(c_{D}(x)) = & ~\frac{3}{2}\vert D\vert -\frac{1}{2}\Psi_{\alpha}(\Delta)\big[\Psi_{\gamma}(\Delta)\Psi_{\beta}(\Delta) + \Psi_{\beta}(\Delta)\Psi_{0}(\Delta) + \Psi_{\beta + \gamma}(\Delta)\Psi_{\beta}(\Delta)\big].
        	\end{split}
        \end{equation}
        From Lemma \ref{U_iLemma}, we have
        \begin{equation}\label{TheMainEqTwo}
        	\Psi_{\alpha}(\Delta)=
        	\begin{cases}
        		2^{\vert L \vert} + 2^{\vert M \vert} - 2^{\vert L \cap M \vert}, \text{  if } \alpha \in \mathfrak{U}_{1};\\
        		2^{\vert L \vert}  - 2^{\vert L \cap M \vert}, \text{  if } \alpha \in \mathfrak{U}_{2};\\
        		2^{\vert M \vert} - 2^{\vert L \cap M \vert}, \text{  if } \alpha \in \mathfrak{U}_{3};\\
        		- 2^{\vert L \cap M \vert}, \text{  if } \alpha \in \mathfrak{U}_{4};\\
        		0, \text{  if } \alpha \in \mathfrak{U}_{5}.\\
        	\end{cases}
        \end{equation}
        Using the above distribution for $\alpha$ and $\beta$, we find the location of $\alpha + \beta$, which is displayed in Table \ref{LocatioOfAlpha_Betal}. For instance, if $\alpha, \beta \in \mathfrak{U}_2$, then
        \begin{equation}
        	\alpha + \beta \in 
        	\begin{cases}
        		\mathfrak{U}_1, \text{ if } \alpha \cap M = \beta \cap M;\\
        		\mathfrak{U}_2, \text{ if } \alpha \cap M \neq \beta \cap M.
        	\end{cases}
        \end{equation}

        We refer [\cite{Wu_Li}, Table VI] for more details about the location of $\alpha + \beta$. 
        \begin{table}[h]
        	\begin{center}
        		\begin{tabular}{|c||c|c|c|c|c|}\hline
        			\diagbox{$\alpha$}{$\alpha + \beta$}{$\beta$}  & $\mathfrak{U}_1$ & $\mathfrak{U}_2$ &  $\mathfrak{U}_3$  & $\mathfrak{U}_4$ & $\mathfrak{U}_5$\\ \hline \hline
        			$\mathfrak{U}_1$  &  $\mathfrak{U}_1$ & $\mathfrak{U}_2$& $\mathfrak{U}_3$ &  $\mathfrak{U}_4$    &  $\mathfrak{U}_5$ \\\hline
        			
        			$\mathfrak{U}_2$ & $\mathfrak{U}_2$& $\mathfrak{U}_1$ or $\mathfrak{U}_2$ &   $\mathfrak{U}_4$   & $\mathfrak{U}_3$  or $\mathfrak{U}_4$ & $\mathfrak{U}_5$   \\ 
        			\hline
        			$\mathfrak{U}_3$   & $\mathfrak{U}_3$ & $\mathfrak{U}_4$&  $\mathfrak{U}_1$ or $\mathfrak{U}_3$  & $\mathfrak{U}_2$ or $\mathfrak{U}_4$& $\mathfrak{U}_5$  \\ \hline
        			
        			$\mathfrak{U}_4$  & $\mathfrak{U}_4$& $\mathfrak{U}_3$ or $\mathfrak{U}_4$ &  $\mathfrak{U}_2$ or $\mathfrak{U}_4$    & $\mathfrak{U}_i (i=1,2 ,3 , 4)$&  $\mathfrak{U}_5$  \\ \hline
        			$\mathfrak{U}_5$ & $\mathfrak{U}_5$ &$\mathfrak{U}_5$ & $\mathfrak{U}_5$ & $\mathfrak{U}_5$ & $\mathfrak{U}_1$ or $\mathfrak{U}_5$ \\ \hline
        		\end{tabular}
        		\caption{Location of $\alpha + \beta$}
        		\label{LocatioOfAlpha_Betal}
        	\end{center}
        \end{table}
        
        Now, by using Lemma \ref{U_iLemma}, Equation \eqref{Eq3.12}, Equation \eqref{TheMainEqTwo} and Table \ref{LocatioOfAlpha_Betal}, we discuss the following cases.
        \begin{enumerate}
        	\item Let $\alpha \in \mathfrak{U}_1.$
        	\begin{itemize}
        		\item If $\beta \in \mathfrak{U}_1, \gamma \in \mathfrak{U}_1$ then $\beta + \gamma \in \mathfrak{U}_1$ and $wt_{Lee}(c_D(x)) = 0$.\\
        		Here in this case, $\#\alpha = 2^{m-\vert L \cup M\vert}, \#\beta =2^{m-\vert L \cup M\vert}, \#\gamma =2^{m-\vert L \cup M\vert}$. Therefore, $\# x = 2^{3m-3\vert L \cup M\vert}$.
        		
        		\item If $\beta \in \mathfrak{U}_1, \gamma \in \mathfrak{U}_2$ then $\beta + \gamma \in \mathfrak{U}_2$ and
        		\begin{equation*}
        			\begin{split}
        				wt_{Lee}(c_D(x)) = &  ~2^{\vert M \vert}(2^{\vert L \vert} + 2^{\vert M \vert} -2^{\vert L \cap M\vert})^2.
        			\end{split}
        		\end{equation*}
        		Here in this case, $\#\alpha = 2^{m-\vert L \cup M\vert}, \#\beta =2^{m-\vert L \cup M\vert}, \#\gamma =(2^{\vert M \setminus L\vert} -1)2^{m-\vert L \cup M\vert}$. Therefore, $\# x = 2^{3m-2\vert L \cup M\vert - \vert L \vert } -2^{3m -3\vert L\cup M\vert }$.
        		
        		\item If $\beta \in \mathfrak{U}_1, \gamma \in \mathfrak{U}_3$ then $\beta + \gamma \in \mathfrak{U}_3$ and
        		\begin{equation*}
        			\begin{split}
        				wt_{Lee}(c_D(x)) = &  ~2^{\vert L \vert}(2^{\vert L \vert} + 2^{\vert M \vert} -2^{\vert L \cap M\vert})^2.
        			\end{split}
        		\end{equation*}
        		Here in this case, $\#\alpha = 2^{m-\vert L \cup M\vert}, \#\beta = 2^{m-\vert L \cup M\vert}, \#\gamma =(2^{\vert L \setminus M\vert} -1)2^{m-\vert L \cup M\vert}$. Therefore, $\# x = 2^{3m-2\vert L \cup M\vert - \vert M \vert } -2^{3m -3\vert L\cup M\vert }$.
        		
        		\item If $\beta \in \mathfrak{U}_1, \gamma \in \mathfrak{U}_4$ then $\beta + \gamma \in \mathfrak{U}_4$ and
        		\begin{equation*}
        			\begin{split}
        				wt_{Lee}(c_D(x)) = &  ~(2^{\vert L \vert} + 2^{\vert M\vert })(2^{\vert L \vert} + 2^{\vert M \vert} -2^{\vert L \cap M\vert})^2.
        			\end{split}
        		\end{equation*}
        		Here in this case, $\#\alpha = 2^{m-\vert L \cup M\vert}, \#\beta =2^{m-\vert L \cup M\vert}, \#\gamma =(2^{\vert L \setminus M\vert} -1)(2^{\vert M \setminus L\vert} -1)2^{m-\vert L \cup M\vert}$. Therefore, $\# x = 2^{3m-2\vert L \cup M\vert - \vert L\cap M \vert } - 2^{3m - 2\vert L \cup M \vert -\vert L \vert} - 2^{3m - 2\vert L \cup M \vert -\vert M \vert}+2^{3m -3\vert L\cup M\vert }$.
        		
        		\item If $\beta \in \mathfrak{U}_1, \gamma \in \mathfrak{U}_5$ then $\beta + \gamma \in \mathfrak{U}_5$ and
        		\begin{equation*}
        			\begin{split}
        				wt_{Lee}(c_D(x)) = &  ~(2^{\vert L \vert} + 2^{\vert M \vert} -2^{\vert L \cap M\vert})^3.
        			\end{split}
        		\end{equation*}
        		Here in this case, $\#\alpha = 2^{m-\vert L \cup M\vert}, \#\beta =2^{m-\vert L \cup M\vert}, \#\gamma =(2^{\vert L \cap M\vert} -1)2^{m-\vert L \cap M\vert}$. Therefore, $\# x = 2^{3m-2\vert L \cup M\vert} - 2^{3m - 2\vert L \cup M \vert -\vert L\cap M \vert} $.
        	\end{itemize}
        \end{enumerate}
        Similarly, one can explore the other cases, namely, $ \beta \in \mathfrak{U}_j, \gamma \in \mathfrak{U}_k, 2\leq j \leq 5, 1\leq k\leq 5$ for $\alpha \in \mathfrak{U}_i, 1\leq i \leq 5$ and obtain the corresponding Lee weights. 

        \section{Gray images and minimal linear codes}\label{section4}
        Here we discuss minimality and self-orthogonality of the binary codes derived from $C_D$ through the Gray map $\Phi$ defined by Equation \eqref{PhiEq}. \par 
        Let $C$ be a linear code over $\mathbb{F}_{2}$. An element $v_{0}\in C\setminus \{0\}$ is called \textit{minimal} if $v\preceq v_{0}$ and $v\in C\setminus \{0\}$ imply $v=v_{0}$. If every codeword of $C\setminus \{0\}$ is minimal then we say that $C$ is a \textit{minimal code}. Now we recall a result from \cite{suffConMinimalcode} that describes a condition for minimality of a linear code over a finite field.    
        \begin{lemma}\label{minimal_lemma}
        	\cite{suffConMinimalcode}\textnormal{(Ashikhmin-Barg)}
        	Consider a linear code $C$ over $\mathbb{F}_q$ with $w_o$ and $w_\infty$ as the minimum and maximum Hamming weights of its nonzero codewords, respectively. If $\frac{w_0}{w_{\infty}}> \frac{q-1}{q}$, then $C$ is minimal.
        \end{lemma}
        Next we recall a sufficient condition (see Theorem $1.4.8 ~(ii)$ of \cite{wchuffman}) for binary linear codes to be self-orthogonal.
        \begin{theorem}\cite{wchuffman}\label{orthogonal_lemma}
          Assume that $C$ is a binary linear code and Hamming weight of every nonzero elements of $C$ is $4k$ for some $k\in \mathbb{N}$. Then $C$ is a self-orthogonal linear code.
        \end{theorem}
        \begin{proposition}\label{section4theoremOrtho}
        	Assume $C_D$ is as in Theorem \ref{main}. Then the Gray image $\Phi(C_D)$ is a binary self-orthogonal code provided $L, M, N $ are nonempty subsets of $[m]$.
        \end{proposition}
        The following result describes the minimality condition for the binary code $\Phi(C_D)$ for each $D$ as in Theorem \ref{main}.
        \begin{theorem}\label{sec4theorem}
        	Let $m\in \mathbb{N}$ and let $L, M, N \subseteq [m]$. Assume $C_D$ is as in Theorem \ref{main}. Let $w_0=\textnormal{min}\{wt_H(c) : c \in \Phi(C_D)\setminus \{0\}\}$ and $w_{\infty}=\textnormal{max}\{wt_H(c) : c \in \Phi(C_D)\setminus \{0\}\}$. Then the sufficient conditions for the Gray image $\Phi(C_D)$ to be a minimal code is given by Table \ref{table:minimalcondition} for each $D$.
        
          \begin{table}[h]
          	\centering
          	  \begin{adjustbox}{width=\textwidth}
          		\begin{tabular}{ c | c | c | c | c  }
          			\hline
          		S.N. &	$D$ as in&  $w_0$   & $w_{\infty}$   & minimality condition \\
          			\hline
          	$1$	&	Theorem \ref{main}(\ref{proposition:1}) & $2^{\vert L\vert +\vert M\vert +\vert N\vert }$ &  $3\times 2^{\vert L\vert +\vert M\vert +\vert N\vert -1 }$   &no conditions\\
          			\hline
          	$2$	&	 Theorem \ref{main}(\ref{proposition:2}) & $(2^m-2^{\vert L\vert})\times 2^{\vert M\vert + \vert N\vert }$ &   $3\times 2^{m+\vert M\vert +\vert N\vert -1}$  &$\vert L\vert \leq m-3$\\
          			\hline
          	$3$	&	 Theorem \ref{main}(\ref{proposition:4}) & $(2^m-2^{\vert M\vert })\times 2^{\vert L\vert +\vert N\vert }$ &  $3\times 2^{m+\vert L\vert +\vert N\vert -1}$   & $\vert M \vert \leq m-3$\\
          			\hline
          	$4$	&     Theorem \ref{main}(\ref{proposition:3}) & $(2^m-2^{\vert N\vert })\times 2^{\vert L\vert +\vert M\vert }$ &  $2^{\vert L\vert +\vert M\vert -1} \times (3 \times 2^m -2^{\vert N \vert +1})$  & $\vert N\vert \leq  m-2$\\
          			\hline
          	\multirow{2}{*}{$5$}	&	\multirow{2}{*}{Theorem \ref{main}(\ref{proposition:5})} & \multirow{2}{*}{$(2^m - 2^{\vert L \vert })(2^m-2^{\vert M\vert })\times 2^{\vert N\vert }$}& $3\times (2^m - 2^{\vert L \vert })\times 2^{m + \vert N\vert -1}$, if $\vert L \vert \leq \vert M\vert $ & $\vert M\vert \leq m-3 $\\
          			\cline{4-5}
          	    & & &$3\times (2^m-2^{\vert M\vert })\times 2^{m+ \vert N\vert -1}$ , if $\vert L \vert \geq \vert M\vert +1$ & $\vert L\vert \leq m-3$ \\
          	    \hline
          	\multirow{2}{*}{$6$}	&	\multirow{2}{*}{Theorem \ref{main}(\ref{proposition:6})} & \multirow{2}{*}{$(2^m - 2^{\vert L \vert })(2^m-2^{\vert N\vert })\times 2^{\vert M\vert }$}& $3\times (2^m - 2^{\vert N \vert })\times 2^{m + \vert M\vert -1}$, if $\vert N \vert \leq \vert L\vert +1$ & $\vert L\vert \leq m-3 $\\
          	\cline{4-5}
          	& & &$(2^m-2^{\vert L\vert })(3\times 2^m-2^{\vert N\vert +1})\times 2^{\vert M\vert -1}$ , if $\vert N \vert \geq \vert L\vert +2 $ and $M\not\subset N$ & $\vert N\vert \leq m-2$ \\
          	\hline
          	    
          	\multirow{2}{*}{$7$}	&	\multirow{2}{*}{Theorem \ref{main}(\ref{proposition:7})} & \multirow{2}{*}{$(2^m - 2^{\vert M \vert })(2^m-2^{\vert N\vert })\times 2^{\vert L\vert }$} &  $(2^m-2^{\vert M\vert })(3\times 2^m -2^{\vert N \vert +1})\times 2^{\vert L\vert -1} + (2^m -2^{\vert N\vert })\times 2^{\vert L \vert +\vert M \vert -1}$ & \multirow{2}{*}{no conditions}\\

          		& &  & $  -2^{\vert L\vert +\vert M\vert +\vert N\vert -1}$, if $\vert M\vert \leq m-2, \vert N\vert \leq m-2$ &\\
          	
          			\hline
          \multirow{3}{*}{$8$}	&	\multirow{3}{*}{Theorem \ref{main}(\ref{proposition:8})} & \multirow{3}{*}{$(2^m - 2^{\vert L \vert })(2^m-2^{\vert M\vert })(2^m-2^{\vert N\vert })$}& $(2^m-2^{\vert L\vert })(2^m-2^{\vert M\vert })(3\times 2^{m-1} -2^{\vert N\vert })$ & \multirow{2}{*}{$\vert M\vert \leq m-2, \vert N\vert \leq m-2 $}\\
          & & & $ + (2^m-2^{\vert L \vert })(2^m-2^{\vert N\vert +1})\times 2^{\vert M \vert -1}$&\\
          \cline{4-5}
          & & &$ 3\times (2^m - 2^{\vert M\vert })(2^m-2^{\vert N\vert })\times 2^{m-1}$ & $\vert L\vert \leq m-3$ \\
          			\hline
          		\end{tabular}
          	\end{adjustbox}
          	\caption{Minimality conditions in Theorem \ref{sec4theorem}}
          	\label{table:minimalcondition}
          \end{table}
        	
        \end{theorem}
        \proof Here we prove S.N. $(6)$ of Table \ref{table:minimalcondition}.\\
        First we consider the case when $\vert N \vert \leq \vert L \vert +1 $.\\
        Then we have $w_0 = (2^m - 2^{\vert L \vert })(2^m-2^{\vert N \vert })\times 2^{\vert M \vert }$ and $w_ {\infty} = 3\times (2^m - 2^{\vert N \vert })\times 2^{m + \vert M\vert -1}$.\\
        Now, $\frac{w_0}{w_{\infty}} = \frac{(2^m - 2^{\vert L \vert })(2^m-2^{\vert N \vert })\times 2^{\vert M \vert } }{3\times (2^m - 2^{\vert N \vert })\times 2^{m + \vert M\vert -1} } > \frac{1}{2} \iff 2^{m-1} > 2^{\vert L \vert +1}$, which is true iff $\vert L \vert \leq m-3$. Therefore, by Lemma \ref{minimal_lemma}, $\Phi(C_D)$ is minimal if $\vert L \vert \leq m-3$.\\
        On the other hand, if $\vert N \vert \geq \vert L \vert +2 $ then $w_0 = (2^m - 2^{\vert L \vert })(2^m-2^{\vert N \vert })\times 2^{\vert M \vert }$ and $w_ {\infty} = (2^m-2^{\vert L\vert })(3\times 2^m-2^{\vert N\vert +1})\times 2^{\vert M\vert -1}$.\\
        Now, $\frac{w_0}{w_{\infty}} = \frac{(2^m - 2^{\vert L \vert })(2^m-2^{\vert N \vert })\times 2^{\vert M \vert } }{(2^m-2^{\vert L\vert })(3\times 2^m-2^{\vert N\vert +1})\times 2^{\vert M\vert -1}} > \frac{1}{2} \iff 2^{m} > 2^{\vert N \vert +1}$, which is true iff $\vert N \vert \leq m-2$. Therefore, by Lemma \ref{minimal_lemma}, $\Phi(C_D)$ is minimal if $\vert N \vert \leq m-2$.\\
        Similarly, one can prove other parts also. \qed
        The following examples illustrate Theorem \ref{main}, Proposition \ref{section4theoremOrtho} and Theorem \ref{sec4theorem}.
        \begin{example}
        	\begin{enumerate}
        		\item Let $m = 5$, $L = \{1, 2\}, M=\{2, 3\}, N=\{3, 4\}$. Then $C_D$ (as in Theorem \ref{main}(\ref{proposition:1})) is a $2$-weight code with Lee weight enumerator $X^{192} + 9 X^{128}Y^{64} + 118 X^{96} Y^{96}$. By Proposition \ref{section4theoremOrtho} and Theorem \ref{sec4theorem}, $\Phi(C_D)$ is a $[192, 7, 64]$ binary minimal and self-orthogonal code.
        		
        		\item Let $m = 5$, $L = \{1, 2, 3\}, M=\{2, 3\}, N=\{3, 4\}$. Then $C_D$ (as in Theorem \ref{main}(\ref{proposition:2})) is a $4$-weight code with Lee weight enumerator $X^{1152} + 9 X^{768}Y^{384} + 984 X^{576} Y^{576} + 27 X^{512}Y^{640} +3X^{384}Y^{768}$. By Proposition \ref{section4theoremOrtho} and Theorem \ref{sec4theorem}, $\Phi(C_D)$ is a $[1152, 10, 384]$ binary minimal and self-orthogonal code.
        		
        		\item Let $m = 5$, $L = \{1, 2, 3\}, M=\{2, 3\}, N=\{2, 3, 4\}$. Then $C_D$ (as in Theorem \ref{main}(\ref{proposition:4})) is a $5$-weight code with Lee weight enumerator $X^{5376} + 24 X^{3584}Y^{1792} + 13 X^{3328} Y^{2048} + 7936 X^{2688} Y^{2688} + 200 X^{2560} Y^{2816} + 18 X^{2304} Y^{3072}$. By Proposition \ref{section4theoremOrtho} and Theorem \ref{sec4theorem}, $\Phi(C_D)$ is a $[5376, 13, 1792]$ binary minimal and self-orthogonal code.
        		\item Let $m = 4$, $L = \{1, 2, 3\}, M=\{1, 2\}, N=\{2, 3\}$. Then $C_D$ (as in Theorem \ref{main}(\ref{proposition:3})) is a $4$-weight code with Lee weight enumerator $X^{1152} + 14 X^{768}Y^{384} + X^{640} Y^{512} +492 X^{576}Y^{576} +4X^{512}Y^{640}$. By Proposition \ref{section4theoremOrtho} and Theorem \ref{sec4theorem}, $\Phi(C_D)$ is a $[1152, 9, 384]$ binary minimal and self-orthogonal code.
        		\item Let $m = 5$, $L = \{1\}, M=\{2, 3\}, N=\{2, 3, 4\}$. Then $C_D$ (as in Theorem \ref{main}(\ref{proposition:5})) is a $10$-weight code with Lee weight enumerator $X^{20160} + 24 X^{13440}Y^{6720} + 13 X^{12480} Y^{7680} + 270 X^{10176}Y^{9984} + 3000 X^{10112}Y^{10048} + 28672 X^{10080} Y^{10080} + 195 X^{9920}Y^{10240}\\ + 360X^{9856}Y^{10304} + 200 X^{9600}Y^{10560} +15 X^{9408}Y^{10752} +18X^{8640} Y^{11520}$. By Proposition \ref{section4theoremOrtho} and Theorem \ref{sec4theorem}, $\Phi(C_D)$ is a $[20160, 15, 6720]$ binary minimal and self-orthogonal code.
        		\item Let $m = 5$, $L = \{1, 2\}, M=\{3\}, N=\{2, 3, 4\}$. Then $C_D$ (as in Theorem \ref{main}(\ref{proposition:5})) is a $10$-weight code with Lee weight enumerator $X^{20160} + 16 X^{13440}Y^{6720} + 21X^{12992} Y^{7168} +294 X^{10176}Y^{9984} + 3024X^{10112}Y^{10048} +28672 X^{10080} Y^{10080} + 147X^{9664}Y^{10496}\\ +112 X^{9600}Y^{10560} + 432 X^{9856} Y^{10304} + 42 X^{9408}Y^{10752} + 7X^{8640}Y^{11520} $. By Proposition \ref{section4theoremOrtho} and Theorem \ref{sec4theorem}, $\Phi(C_D)$ is a $[20160, 15, 6720]$ binary minimal and self-orthogonal code.
        		\item Let $m = 4$, $L = \{1\}, M=\{2, 3\}, N=\{1, 3\}$. Then $C_D$ (as in Theorem \ref{main}(\ref{proposition:6})) is a $8$-weight code with Lee weight enumerator $X^{2016} + 56 X^{1344}Y^{672} + 112 X^{1024} Y^{992} + 3456 X^{1008}Y^{1008} + 28 X^{992}Y^{1024} + 392 X^{960}Y^{1056}  + 4X^{1120}Y^{896} + 16 X^{896}Y^{1120}\\ + 28X^{864}Y^{1152}$. By Proposition \ref{section4theoremOrtho} and Theorem \ref{sec4theorem}, $\Phi(C_D)$ is a $[2016, 10, 672]$ binary minimal and self-orthogonal code.
        		
        	\end{enumerate}
        \end{example}
        \section{Binary subfield-like codes from linear codes over $\mathcal{R}$}\label{section5}
        
        This section constructs binary linear codes from $C_{D}$ which are similar to binary subfield codes studied in \cite{Ding_Heng, Wu_Li}.\\
        We consider the map $\tau : \mathcal{R} \longrightarrow \mathbb{F}_2$ defined by 
        \begin{equation}
        	\tau(\alpha + u\beta + u^2 \gamma) = \alpha.
        \end{equation}
        This extends to a map from $\mathcal{R}^m$ to $\mathbb{F}_2^{m}$ component-wise for any $m \in \mathbb{N}$. Note that $\tau: C_{D} \longrightarrow \tau(C_{D})\subseteq \mathbb{F}_{2}^{\vert D\vert}$ defined by
        \begin{equation}\label{tauMap}
        	\tau\big((x\cdot d)_{d\in D}\big) = \big(\tau(x\cdot d)\big)_{d\in D}
        \end{equation}
        is a surjective linear transformation of $\mathbb{F}_2$-vector spaces and $\tau(C_{D})$ is a binary linear code of length $\vert D\vert$.
        \begin{remark}
        	Let $\mathcal{B}=\{ 1 + u^2 <  u^2 <  u + u^2\}$. This is an ordered basis of the vector space $\mathcal{R}$ over $\mathbb{F}_2$. For $x \in \mathcal{R}$, let $l_x: \mathcal{R} \longrightarrow \mathcal{R}$ be the left multiplication map, that is,
        	\begin{equation}
        		l_x(y) = xy. 
        	\end{equation}
        	Then $l_x$ is an $\mathbb{F}_2$-linear transformation. Define the map $\eta : \mathcal{R}  \longrightarrow M_{3\times 3}(\mathbb{F}_2)$ by $\eta(x) = [l_x]_{\mathcal{B}}$, the matrix representation of $l_x$ with respect to $\mathcal{B}$. Then,
        	\begin{scriptsize}
        	\begin{align*}
        		\eta(0) &= \begin{pmatrix}
        			0 & 0 & 0\\
        			0 & 0 & 0\\
        			0 & 0 & 0
        		\end{pmatrix},&
        		\eta(1) &= \begin{pmatrix}
        			1 & 0 & 0\\
        			0 & 1 & 0\\
        			0 & 0 & 1
        		\end{pmatrix}, &
        		\eta(u)& = \begin{pmatrix}
        			0 & 0 & 0\\
        			0 & 1 & 0\\
        			0 & 1 & 1
        		\end{pmatrix},&
        		\eta(u^2) &=\begin{pmatrix}
        			0 & 0 & 0\\
        			0 & 1 & 0\\
        			0 & 0 & 1
        		\end{pmatrix},\\
        		\eta(1 + u) &=\begin{pmatrix}
        			1 & 0 & 0\\
        			0 & 0 & 0\\
        			0 & 1 & 0
        		\end{pmatrix},&
        		\eta(1 + u^2)&=\begin{pmatrix}
        			1 & 0 & 0\\
        			0 & 0 & 0\\
        			0 & 0 & 0
        		\end{pmatrix},&
        		\eta(u + u^2)&=\begin{pmatrix}
        			0 & 0 & 0\\
        			0 & 0 & 0\\
        			0 & 1 & 0
        		\end{pmatrix},&
        		\eta(1+ u + u^2)&=\begin{pmatrix}
        			1 & 0 & 0\\
        			0 & 1 & 0\\
        			0 & 1 & 1
        		\end{pmatrix}.
        	\end{align*}
            \end{scriptsize}

        	Define the $\mathbb{F}_2$-linear map $Tr : \mathcal{R} \longrightarrow \mathbb{F}_2$ by
           \begin{equation}\label{traceMap}
        		Tr(x): = \textnormal{the trace of the matrix } \eta(x).
        	\end{equation}
        	Then, $Tr(\alpha + u\beta + u^2\gamma ) = \alpha$. Since the $Tr$ map is degenerate, that is, $Tr(xy) = 0$ $\forall$ $x \in \mathcal{R} \nRightarrow y = 0$, the analogue (for $\mathcal{R}$) of Lemma $2.2$ and Lemma $2.3$ in \cite{Ding_Heng} are not true.
        \end{remark}
        
           Let $x = (1+u^2)\alpha + u^2\beta + (u + u^2)\gamma \in \mathcal{R}^m$ and $D = (1+u^2)D_1 + u^2D_2 + (u + u^2)D_3 \subseteq \mathcal{R}^m$. By Equation \eqref{c_DMap}, we have
           \begin{equation}
           	\tau(C_{D}) = \{\tau(c_{D}(x)) = (\tau(x\cdot d))_{d\in D} : x \in \mathcal{R}^m \}\subseteq \mathbb{F}_{2}^{\vert D\vert}.
           \end{equation}
           Note that $\tau(C_{D})$ is a binary linear code. 
           Define $T := \tau \circ c_{D}$, where $c_{D}$ is defined by Equation \eqref{c_DMap} and $\tau$ is defined by Equation \eqref{tauMap}.
           Observe that the map $T: \mathcal{R}^m \longrightarrow C_{D} \longrightarrow \tau(C_D) \subseteq \mathbb{F}_{2}^{\vert D \vert}$ defined by 
           \begin{equation}
           	T(x) = \big(\tau(c_{D}(x))\big) = \big(\tau(x\cdot d)\big)_{d\in D}
           \end{equation}
           is a surjective linear transformation of $\mathbb{F}_2$-vector spaces. By the first isomorphism theorem, we have $\vert \tau(C_{D})\vert = \frac{\vert \mathcal{R}^m \vert}{\vert \ker(T)\vert}$.\\
           Observe that given any nonzero $\mathbb{F}_{2}$-functional $\sigma$ of $\mathcal{R}$, analogously one can construct the binary code $\sigma(C_{D})$. We refer to $\sigma(C_{D})$ as the \textit{subfield-like} code of $C_{D}$ corresponding to $\sigma$. In this section, we study the subfield-like code $\tau(C_{D})$. The definition of subfield-like code is motivated  by the description of subfield codes in \cite{Ding_Heng, Wu_Li}.\\
           Now, we find the Hamming weight of an element of $\tau(C_{D})$.
           \begin{equation}\label{KeyEq_SF_like}
           	\begin{split}
           		wt_{H}\big(\tau(c_D(x))\big) 
           		 = & ~wt_{H}\big((\tau(((1+u^2)\alpha + u^2\beta + (u+u^2)\gamma)\cdot ((1+u^2)d_1 \\ 
           		& + u^2d_2 + (u+u^2)d_3) ))_{d_1 \in D_1, d_2 \in D_2, d_3 \in D_3}\big)\\
           		= & ~wt_{H}\big((\alpha d_1 )_{d_1 \in D_1, d_2 \in D_2, d_3 \in D_3}\big)\\
           		= & ~\vert D \vert - \frac{1}{2}\sum_{d_1\in D_1}\sum_{d_2\in D_2}\sum_{d_3\in D_3}\big(1 + (-1)^{\alpha d_1}\big)\\
           		= & ~\frac{1}{2}\vert D\vert - \frac{1}{2}\sum_{d_1\in D_1}(-1)^{\alpha d_1}\sum_{d_2\in D_2}(1)\sum_{d_3\in D_3}(1).
           	\end{split}
           \end{equation}
    \subsection{Using simplicial complexes having one maximal element}
           In this subsection, we study the subfield-like code $\tau(C_D)$ from the linear code $C_D$ over $\mathcal{R}$, where the defining set $D$ is constructed using simplicial complexes having one maximal element.
        \begin{theorem}\label{subfield_main}
        	Suppose $m\in \mathbb{N}$ and $L, M, N \subseteq [m]$.
        	\begin{enumerate}
        		\item \label{item_1}
        		Let $D = (1+u^2)\Delta_{L} + u^2\Delta_M + (u+u^2)\Delta_N \subseteq \mathcal{R}^m$. Then $\tau(C_D)$ is a binary $[2^{\vert L \vert + \vert M \vert + \vert N \vert }, \vert L \vert,\\ 2^{\vert L \vert +\vert M \vert + \vert N \vert -1}]$ linear $1$-weight code and its weight distribution is displayed in Table \ref{subfield_1}. In particular, $\tau(C_D)$ is a minimal code. In addition, it is self-orthogonal provided $\vert L \vert +\vert M \vert + \vert N \vert \geq 3$. Suppose $Z_i = \vert \{x \in \mathcal{R}^m : wt_H(\tau(c_D(x))) = i\} \vert$, where $i=0, 2^{\vert L \vert + \vert M \vert + \vert N \vert -1}$. Then
        		\begin{center}
        			$Z_0 = 2^{3m-\vert L \vert}$ and $Z_j=2^{3m-\vert L \vert }\times (2^{\vert L \vert} -1)$,
        		\end{center}
        		where $j= 2^{\vert L \vert + \vert M \vert+ \vert N \vert -1}$.
        		\begin{table}[]
        			\centering
        			\scalebox{0.9}{\begin{tabular}{  c | c  }
        				\hline
        				Hamming weight   & Number of codewords \\
        				\hline
        				$2^{\vert L \vert + \vert M \vert + \vert N \vert-1 }$ & $2^{\vert L \vert} -1$\\        				
        				\hline
        				$0$ & $1$\\
        				\hline
        			\end{tabular}}
        			\caption{Hamming weight distribution in Theorem \ref{subfield_main}(\ref{item_1})}
        			\label{subfield_1}	
        		\end{table}
        	\item \label{item_2}
        	Let $D = (1+u^2)\Delta_{L}^{\textnormal{c}} + u^2\Delta_M + (u+u^2)\Delta_N \subseteq \mathcal{R}^m$. Then $\tau(C_D)$ is a binary $[(2^m - 2^{\vert L \vert})2^{\vert M \vert + \vert N \vert }, m, (2^m - 2^{\vert L \vert})2^{\vert M \vert + \vert N \vert -1}]$ linear $2$-weight code and its weight distribution is displayed in Table \ref{subfield_2}. If $\vert L \vert \leq m-2$ then $\tau(C_D)$ is minimal. In addition, it is self-orthogonal provided $\vert L \vert +\vert M \vert + \vert N \vert \geq 3$. Suppose $Z_i = \vert \{x \in \mathcal{R}^m : wt_H(\tau(c_D(x))) = i\} \vert$, where $i=0, (2^m - 2^{\vert L \vert})2^{\vert M \vert + \vert N \vert -1}, 2^{m + \vert M \vert + \vert N \vert -1}$. Then
        	\begin{center}
        		$Z_0 = 2^{2m}$, $Z_{j_1}=2^{3m-\vert L \vert }\times (2^{\vert L \vert} -1)$ and $Z_{j_2}=2^{2m}\times (2^{m-\vert L \vert} -1)$,
        	\end{center}
        	where $j_1= (2^m - 2^{\vert L \vert})2^{\vert M \vert + \vert N \vert -1}, j_2 = 2^{m + \vert M \vert + \vert N \vert -1}$.
        	\begin{table}[]
        		\centering
        		\scalebox{0.8}{\begin{tabular}{  c | c  }
        			\hline
        			Hamming weight   & Number of codewords \\
        			\hline
        			$2^{m + \vert M \vert + \vert N \vert -1}$ & $2^{m-\vert L \vert} -1$\\
        			\hline
        			$(2^m - 2^{\vert L \vert})2^{\vert M \vert + \vert N \vert -1}$ & $2^{m-\vert L \vert }\times (2^{\vert L \vert} -1)$\\        				
        			\hline
        			$0$ & $1$\\
        			\hline
        		\end{tabular}}
        		\caption{Hamming weight distribution in Theorem \ref{subfield_main}(\ref{item_2})}
        		\label{subfield_2}	
        	\end{table}
            \begin{enumerate}
            	\item \label{item_2.1}
            	Let $1 \leq \vert L \vert +\vert M \vert + \vert N \vert \leq m $ and $\theta_1 = 2^{\vert M \vert +\vert N \vert} -1 $. If $0 \leq \theta_1 < \vert L \vert +\vert M \vert + \vert N \vert $ then $\tau(C_D)$ is optimal with respect to the Griesmer bound.
            	
            	\item \label{item_2.2}
            	Let $m < \vert L \vert +\vert M \vert + \vert N \vert \leq 3m -1 $ and $\theta_2 = 2^{\vert L \vert +\vert M \vert + \vert N \vert -m}(2^{m- \vert L \vert} -1) $. If $0 < \theta_2 < m $ then $\tau(C_D)$ is optimal with respect to the Griesmer bound.
            \end{enumerate}
            \item \label{item_3}
             Let $D = (1+u^2)\Delta_{L} + u^2\Delta_M^{\textnormal{c}} + (u+u^2)\Delta_N \subseteq \mathcal{R}^m$. Then $\tau(C_D)$ is a binary $[(2^m - 2^{\vert M \vert})2^{\vert L \vert + \vert N \vert }, \vert L \vert, (2^m - 2^{\vert M \vert})2^{\vert L \vert + \vert N \vert -1}]$ linear $1$-weight code and its weight distribution is displayed in Table \ref{subfield_3}. In particular, it is minimal. In addition, it is self-orthogonal provided $\vert L \vert +\vert M \vert + \vert N \vert \geq 3$. Suppose $Z_i = \vert \{x \in \mathcal{R}^m : wt_H(\tau(c_D(x))) = i\} \vert$, where $i=0, (2^m - 2^{\vert M \vert})2^{\vert L \vert + \vert N \vert -1}$. Then
             \begin{center}
             	$Z_0 = 2^{3m-\vert L \vert }$, $Z_{j}=2^{3m-\vert L \vert }\times (2^{\vert L \vert} -1)$,
             \end{center}
             where $j= (2^m - 2^{\vert M \vert})2^{\vert L \vert + \vert N \vert -1}$.
             \begin{table}[]
             	\centering
             	\scalebox{0.9}{\begin{tabular}{  c | c  }
             		\hline
             		Hamming weight   & Number of codewords \\
             		\hline
             		$(2^m - 2^{\vert M \vert})2^{\vert L \vert + \vert N \vert -1}$ & $2^{\vert L \vert} -1$\\        				
             		\hline
             		$0$ & $1$\\
             		\hline
             	\end{tabular}}
             	\caption{Hamming weight distribution in Theorem \ref{subfield_main}(\ref{item_3})}
             	\label{subfield_3}	
             \end{table}
             \item \label{item_4}
             Let $D = (1+u^2)\Delta_{L} + u^2\Delta_M + (u+u^2)\Delta_N^{\textnormal{c}} \subseteq \mathcal{R}^m$. Then $\tau(C_D)$ is a binary $[(2^m - 2^{\vert N \vert})2^{\vert L \vert + \vert M \vert }, \vert L \vert, (2^m - 2^{\vert N \vert})2^{\vert L \vert + \vert M \vert -1}]$ linear $1$-weight code and its weight distribution is displayed in Table \ref{subfield_4}. In particular, it is minimal. In addition, it is self-orthogonal provided $\vert L \vert +\vert M \vert + \vert N \vert \geq 3$. Suppose $Z_i = \vert \{x \in \mathcal{R}^m : wt_H(\tau(c_D(x))) = i\} \vert$, where $i=0, (2^m - 2^{\vert N \vert})2^{\vert L \vert + \vert M \vert -1}$. Then
             \begin{center}
             	$Z_0 = 2^{3m-\vert L \vert }$, $Z_{j}=2^{3m-\vert L \vert }\times (2^{\vert L \vert} -1)$,
             \end{center}
             where $j= (2^m - 2^{\vert N \vert})2^{\vert L \vert + \vert M \vert -1}$.
             \begin{table}[]
             	\centering
             	\scalebox{0.9}{\begin{tabular}{  c | c  }
             		\hline
             		Hamming weight   & Number of codewords \\
             		\hline
             		$(2^m - 2^{\vert N \vert})2^{\vert L \vert + \vert M \vert -1}$ & $2^{\vert L \vert} -1$\\        				
             		\hline
             		$0$ & $1$\\
             		\hline
             	\end{tabular}}
             	\caption{Hamming weight distribution in Theorem \ref{subfield_main}(\ref{item_4})}
             	\label{subfield_4}	
             \end{table}
             \item \label{item_5}
             Let $D = (1+u^2)\Delta_{L}^{\textnormal{c}} + u^2\Delta_M^{\textnormal{c}} + (u+u^2)\Delta_N \subseteq \mathcal{R}^m$. Then $\tau(C_D)$ is a binary $[(2^m - 2^{\vert L \vert})(2^m - 2^{\vert M \vert})2^{\vert N \vert }, m, (2^m - 2^{\vert L \vert})(2^m - 2^{\vert M \vert})2^{\vert N \vert -1}]$ linear $2$-weight code and its weight distribution is displayed in Table \ref{subfield_5}. If $\vert L \vert \leq m-2$ then $\tau(C_D)$ is minimal. In addition, it is self-orthogonal provided $\vert L \vert +\vert M \vert + \vert N \vert \geq 3$. Suppose $Z_i = \vert \{x \in \mathcal{R}^m : wt_H(\tau(c_D(x))) = i\} \vert$, where $i=0, (2^m - 2^{\vert L \vert})(2^m - 2^{\vert M \vert})2^{\vert N \vert -1}, (2^m - 2^{\vert M \vert})2^{m+\vert N \vert -1}$. Then
             \begin{center}
             	$Z_0 = 2^{2m}$, $Z_{j_1}=2^{3m-\vert L \vert }\times (2^{\vert L \vert} -1)$ and $Z_{j_2}=2^{2m}\times (2^{m-\vert L \vert} -1)$,
             \end{center}
             where $j_1= (2^m - 2^{\vert L \vert})(2^m - 2^{\vert M \vert})2^{\vert N \vert -1}, j_2 = (2^m - 2^{\vert M \vert})2^{m+\vert N \vert -1}$.
             \begin{table}[]
             	\centering
             	\scalebox{0.8}{\begin{tabular}{  c | c  }
             		\hline
             		Hamming weight   & Number of codewords \\
             		\hline
             		$(2^m - 2^{\vert M \vert})2^{m+\vert N \vert -1}$ & $2^{m-\vert L \vert} -1$\\
             		\hline
             		$(2^m - 2^{\vert L \vert})(2^m - 2^{\vert M \vert})2^{\vert N \vert -1}$ & $2^{m-\vert L \vert }\times (2^{\vert L \vert} -1)$\\        				
             		\hline
             		$0$ & $1$\\
             		\hline
             	\end{tabular}}
             	\caption{Hamming weight distribution in Theorem \ref{subfield_main}(\ref{item_5})}
             	\label{subfield_5}	
             \end{table}
             \item \label{item_6}
             Let $D = (1+u^2)\Delta_{L}^{\textnormal{c}} + u^2\Delta_M + (u+u^2)\Delta_N^{\textnormal{c}} \subseteq \mathcal{R}^m$. Then $\tau(C_D)$ is a binary $[(2^m - 2^{\vert L \vert})(2^m - 2^{\vert N \vert})2^{\vert M \vert }, m, (2^m - 2^{\vert L \vert})(2^m - 2^{\vert N \vert})2^{\vert M \vert -1}]$ linear $2$-weight code and its weight distribution is displayed in Table \ref{subfield_6}. If $\vert L \vert \leq m-2$ then $\tau(C_D)$ is minimal. In addition, it is self-orthogonal provided $\vert L \vert +\vert M \vert + \vert N \vert \geq 3$. Suppose $Z_i = \vert \{x \in \mathcal{R}^m : wt_H(\tau(c_D(x))) = i\} \vert$, where $i=0, (2^m - 2^{\vert L \vert})(2^m - 2^{\vert N \vert})2^{\vert M \vert -1}, (2^m - 2^{\vert N \vert})2^{m+\vert M \vert -1}$. Then
             \begin{center}
             	$Z_0 = 2^{2m}$, $Z_{j_1}=2^{3m-\vert L \vert}(2^{\vert L \vert} -1)$ and $Z_{j_2}=2^{2m}(2^{m-\vert L\vert} -1)$,
             \end{center}
             where $j_1= (2^m - 2^{\vert L \vert})(2^m - 2^{\vert N \vert})2^{\vert M \vert -1}, j_2 = (2^m - 2^{\vert N \vert})2^{m+\vert M \vert -1}$.
             \begin{table}[]
             	\centering
             	\scalebox{0.8}{\begin{tabular}{  c | c  }
             		\hline
             		Hamming weight   & Number of codewords \\
             		\hline
             		$(2^m - 2^{\vert N \vert})2^{m+\vert M \vert -1}$ & $2^{m-\vert L \vert} -1$\\
             		\hline
             		$(2^m - 2^{\vert L \vert})(2^m - 2^{\vert N \vert})2^{\vert M \vert -1}$ & $2^{m-\vert L \vert }\times (2^{\vert L \vert} -1)$\\        				
             		\hline
             		$0$ & $1$\\
             		\hline
             	\end{tabular}}
             	\caption{Hamming weight distribution in Theorem \ref{subfield_main}(\ref{item_6})}
             	\label{subfield_6}	
             \end{table}
             \item \label{item_7}
             Let $D = (1+u^2)\Delta_{L} + u^2\Delta_M^{\textnormal{c}} + (u+u^2)\Delta_N^{\textnormal{c}} \subseteq \mathcal{R}^m$. Then $\tau(C_D)$ is a binary $[(2^m - 2^{\vert M \vert})(2^m - 2^{\vert N \vert})2^{\vert L \vert }, \vert L \vert, (2^m - 2^{\vert M \vert})(2^m - 2^{\vert N \vert})2^{\vert L \vert -1}]$ linear $1$-weight code and its weight distribution is displayed in Table \ref{subfield_7}. In particular, it is minimal. In addition, it is self-orthogonal provided $\vert L \vert +\vert M \vert + \vert N \vert \geq 3$. Suppose $Z_i = \vert \{x \in \mathcal{R}^m : wt_H(\tau(c_D(x))) = i\} \vert$, where $i=0, (2^m - 2^{\vert M \vert})(2^m - 2^{\vert N \vert})2^{\vert L \vert -1}$. Then
             \begin{center}
             	$Z_0 = 2^{3m -\vert L \vert}$, $Z_{j}=2^{3m -\vert L \vert}\times (2^{\vert L \vert} - 1)$,
             \end{center}
             where $j= (2^m - 2^{\vert M \vert})(2^m - 2^{\vert N \vert})2^{\vert L \vert -1}$.
             \begin{table}[]
             	\centering
             	\scalebox{0.9}{\begin{tabular}{  c | c  }
             		\hline
             		Hamming weight   & Number of codewords \\             		
             		\hline
             		$(2^m - 2^{\vert M \vert})(2^m - 2^{\vert N \vert})2^{\vert L \vert -1}$ & $2^{\vert L \vert} -1$\\        				
             		\hline
             		$0$ & $1$\\
             		\hline
             	\end{tabular}}
             	\caption{Hamming weight distribution in Theorem \ref{subfield_main}(\ref{item_7})}
             	\label{subfield_7}	
             \end{table}
             \item \label{item_8}
             Let $D = (1+u^2)\Delta_{L}^{\textnormal{c}} + u^2\Delta_M^{\textnormal{c}} + (u+u^2)\Delta_N^{\textnormal{c}} \subseteq \mathcal{R}^m$. Then $\tau(C_D)$ is a binary $[(2^m - 2^{\vert L \vert})(2^m - 2^{\vert M \vert})(2^m - 2^{\vert N \vert}), m, (2^m - 2^{\vert L \vert})(2^m - 2^{\vert M \vert})(2^{m-1} - 2^{\vert N \vert -1})]$ linear $2$-weight code and its weight distribution is displayed in Table \ref{subfield_8}. If $\vert L \vert \leq m-2$ then $\tau(C_D)$ is minimal. In addition, it is self-orthogonal provided $\vert L \vert +\vert M \vert + \vert N \vert \geq 3$. Suppose $Z_i = \vert \{x \in \mathcal{R}^m : wt_H(\tau(c_D(x))) = i\} \vert$, where $i=0, (2^m - 2^{\vert L \vert})(2^m - 2^{\vert M \vert})(2^{m-1}- 2^{\vert N \vert -1}), 2^{m-1}\times(2^m - 2^{\vert M \vert})(2^{m}- 2^{\vert N \vert })$. Then
             \begin{center}
             	$Z_0 = 2^{2m}$, $Z_{j_1}=2^{3m - \vert L \vert}\times(2^{\vert L \vert} - 1)$ and $Z_{j_2}=2^{2m}\times (2^{m-\vert L \vert} -1)$,
             \end{center}
             where $j_1= (2^m - 2^{\vert L \vert})(2^m - 2^{\vert M \vert})(2^{m-1} - 2^{\vert N \vert -1}), j_2 = 2^{m-1}\times(2^m - 2^{\vert M \vert})(2^{m} - 2^{\vert N \vert })$.
             \begin{table}[]
             	\centering
             	\scalebox{0.8}{\begin{tabular}{  c | c  }
             		\hline
             		Hamming weight   & Number of codewords \\
             		\hline
             		$2^{m-1}\times(2^m - 2^{\vert M \vert})(2^{m}- 2^{\vert N \vert })$ & $2^{m-\vert L \vert} -1$\\
             		\hline
             		$(2^m - 2^{\vert L \vert})(2^m - 2^{\vert M \vert})(2^{m-1}- 2^{\vert N \vert -1})$ & $2^{m-\vert L \vert }\times (2^{\vert L \vert} -1)$\\        				
             		\hline
             		$0$ & $1$\\
             		\hline
             	\end{tabular}}
             	\caption{Hamming weight distribution in Theorem \ref{subfield_main}(\ref{item_8})}
             	\label{subfield_8}	
             \end{table}
        	\end{enumerate}        	
        \end{theorem}
        \proof We prove part (\ref{item_2}) and other parts can be proved similarly.\\
        Observe, the length of $\tau(C_{D})$ is $\vert D\vert = (2^m - 2^{\vert L\vert})2^{\vert M\vert+\vert N \vert}$.
        Suppose $x = (1+u^2)\alpha + u^2 \beta + (u+u^2)\gamma \in \mathcal{R}^m $ and $D = (1+u^2)\Delta_{L}^{\textnormal{c}} + u^2\Delta_M + (u + u^2)\Delta_N $. Then by Equation \eqref{keyeq2}, Equation \eqref{KeyEq_SF_like} and Lemma \ref{Complement_sum}, we have
        \begin{equation*}
        	\begin{split}
        		wt_{H}\big(\tau(c_D(x))\big)  
        		= & \frac{1}{2}\vert D\vert - \frac{1}{2}\sum_{d_1\in \Delta_{L}^{\textnormal{c}}}(-1)^{\alpha d_1}\sum_{d_2\in \Delta_{M}}(1)\sum_{d_3\in \Delta_N}(1)\\
        		 = & (2^m-2^{\vert L \vert })2^{\vert M \vert + \vert N \vert -1} - \big(2^m\delta_{0, \alpha} - 2^{\vert L \vert}\chi(\alpha \vert L)\big)2^{\vert M \vert + \vert N \vert -1}.
        	\end{split}
        \end{equation*}
        Here we discuss the following cases.
        \begin{enumerate}
        	\item If $\alpha =0$ then $wt_{H}\big(\tau(c_D(x))\big) =0$.\\
        	In this case, $\# \alpha =1 , \# \beta =2^m, \# \gamma =2^m$. Therefore, $Z_0 = 2^{2m}$.
        	\item If $\alpha \neq 0$ then $wt_{H}\big(\tau(c_D(x))\big) = (2^m-2^{\vert L \vert })2^{\vert M \vert + \vert N \vert -1} + 2^{\vert L \vert +\vert M \vert + \vert N \vert -1} \chi(\alpha \vert L)$.
        	\begin{enumerate}
        		\item If $\chi(\alpha \vert L) = 0$ then $wt_{H}\big(\tau(c_D(x))\big) = (2^m-2^{\vert L \vert })2^{\vert M \vert + \vert N \vert -1} = j_1$.
        		In this case, by using Lemma \ref{countingLemma}, we get\\
        		$\# \alpha =(2^{\vert L \vert} -1)\times 2^{m-\vert L \vert}$,
        		$\# \beta = 2^m$,
        		$\# \gamma =2^m$.\\
        		Therefore, $Z_{j_1} = (2^{\vert L \vert} -1)\times 2^{3m-\vert L \vert}$.
        		\item If $\chi(\alpha \vert L) = 1$ then $wt_{H}\big(\tau(c_D(x))\big) = 2^{m + \vert M \vert + \vert N \vert -1} =j_2$.
        		In this case, by using Lemma \ref{countingLemma}, we get\\
        		$\# \alpha =(2^{m-\vert L \vert} -1)$,
        		$\# \beta = 2^m$,
        		$\# \gamma =2^m$.\\
        		Therefore, $Z_{j_2} =2^{2m}(2^{m-\vert L \vert} -1) $.
        	\end{enumerate}
        \end{enumerate}
         Note that $T: \mathcal{R}^m \longrightarrow \tau(C_{D}) \subseteq \mathbb{F}_{2}^{\vert D\vert }$ is a surjective homomorphism and $\vert \ker(T) \vert = \vert \{x \in \mathcal{R}^m : \tau(x\cdot d) = 0 ~ \forall ~d \in D\} \vert = Z_0 = 2^{2m}$. By the first isomorphism theorem, we have $\vert \tau(C_{D}) \vert = \frac{\vert \mathcal{R}^m \vert}{\vert \ker(T) \vert} = 2^m$. Hence, $\dim(\tau(C_{D})) = m$.\\
         Note that, the minimum and maximum Hamming weight of nonzero codewords of $\tau(C_{D})$ are $w_0 = (2^m-2^{\vert L \vert })2^{\vert M \vert + \vert N \vert -1}$ and $w_{\infty} = 2^{m + \vert M \vert + \vert N \vert -1}$, respectively. Now we get
         \begin{center}
         	$\frac{w_0}{w_{\infty}} = \frac{(2^m-2^{\vert L \vert })2^{\vert M \vert + \vert N \vert -1}}{2^{m + \vert M \vert + \vert N \vert -1}} > \frac{1}{2} \iff \vert L \vert \leq m-2$.
         \end{center}
         Hence, by Lemma \ref{minimal_lemma}, $\tau(C_{D})$ is minimal provided $\vert L \vert \leq m-2$.\\
         (\ref{item_2.1}) Let $1 \leq \vert L \vert + \vert M \vert + \vert N \vert \leq m$ and $\theta_1 = 2^{\vert M \vert + \vert N \vert } -1$. The parameters of the code $\tau(C_{D})$ are $n = (2^m - 2^{\vert L \vert})2^{\vert M \vert + \vert N \vert}$, $k=m$ and $d=(2^m-2^{\vert L \vert })2^{\vert M \vert + \vert N \vert -1}$. We get,
         \begin{equation*}
         	\begin{split}
         		\sum_{i=0}^{m-1}\left\lceil \frac{2^{m+\vert M \vert + \vert N \vert -1} - 2^{\vert L \vert+\vert M \vert + \vert N \vert -1}}{2^i}\right\rceil = &
         		\sum_{i=0}^{m-1} \frac{2^{m+\vert M \vert + \vert N \vert -1}}{2^i}-
         		\sum_{i=0}^{\vert L \vert + \vert M \vert + \vert N \vert-1}\left\lfloor \frac{ 2^{\vert L \vert+\vert M \vert + \vert N \vert -1}}{2^i}\right\rfloor - \\
         		&~~ \sum_{i=\vert L \vert + \vert M \vert + \vert N \vert}^{m-1}\left\lfloor \frac{ 2^{\vert L \vert+\vert M \vert + \vert N \vert -1}}{2^i}\right\rfloor\\
         		 = & ~(2^{m+ \vert M \vert + \vert N \vert} - 2^{\vert M \vert + \vert N \vert})-(2^{\vert L \vert + \vert M \vert + \vert N \vert} -1) - 0\\
         		 = & ~n -\theta_1.
         	\end{split}
         \end{equation*}
       Therefore, $\sum\limits_{i=0}^{k-1}\lceil \frac{d}{2^i}\rceil \leq n \iff \theta_1 \geq 0$.\\
       On the other hand,\\
       \begin{equation*}
       	\begin{split}
       		\sum_{i=0}^{m-1}\left\lceil \frac{2^{m+\vert M \vert + \vert N \vert -1} - 2^{\vert L \vert+\vert M \vert + \vert N \vert -1} +1 }{2^i}\right\rceil  = &
       		\sum_{i=0}^{m-1} \frac{2^{m+\vert M \vert + \vert N \vert -1}}{2^i}\\
       		&~~
       		-\sum_{i=0}^{\vert L \vert + \vert M \vert + \vert N \vert-1}\left\lfloor \frac{ 2^{\vert L \vert+\vert M \vert + \vert N \vert -1}}{2^i} -\frac{1}{2^i}\right\rfloor \\
       		&~~ -\sum_{i=\vert L \vert + \vert M \vert + \vert N \vert}^{m-1}\left\lfloor \frac{ 2^{\vert L \vert+\vert M \vert + \vert N \vert -1}}{2^i}-\frac{1}{2^i}\right\rfloor\\
       		 = & ~(2^{m+ \vert M \vert + \vert N \vert} - 2^{\vert M \vert + \vert N \vert})-\{(2^{\vert L \vert + \vert M \vert + \vert N \vert} -1) \\ 
       		 & ~~ -(\vert L \vert + \vert M \vert + \vert N \vert)\} - 0\\
       		 = & ~n -\theta_1 + \vert L \vert + \vert M \vert + \vert N \vert.
       	\end{split}
       \end{equation*}
        Therefore, $\sum\limits_{i=0}^{k-1}\lceil \frac{d+1}{2^i}\rceil > n \iff \theta_1 < \vert L \vert + \vert M \vert + \vert N \vert $.\\
        This completes the proof of (\ref{item_2.1}).\\
        Part (\ref{item_2.2}) can be proved using the same argument.\qed
        We give examples to illustrate the above result.
         \begin{example}
         	\begin{enumerate}
         		\item Let $m=5$, $L=\{1, 2, 3, 4\}$, $M=\{2\}$, $N=\{4, 5\}$. If $D$ is as in Theorem \ref{subfield_main}(\ref{item_1}) then $\tau(C_{D})$ is a binary $[128, 4, 64]$-linear code with the Hamming weight enumerator $X^{128} +15 X^{64}Y^{64}$. It is minimal and self-orthogonal.
         		\item \begin{enumerate}
         			   \item Let $m=6$, $L=\{1, 2\}$, $M=\{2\}$, $N=\{5\}$. If $D$ is as in Theorem \ref{subfield_main}(\ref{item_2.1}) then $\tau(C_{D})$ is a binary $[240, 6, 120]$-linear code with the Hamming weight enumerator $X^{240} +48 X^{120}Y^{120} + 15X^{112}Y^{128}$. Note that $\tau(C_{D})$ is optimal according to the Database \cite{BoundTable}. It is minimal and self-orthogonal.
         			   \item Let $m=5$, $L=\{1, 2, 3, 4\}$, $M=\{2, 3\}$, $N=\{5\}$. If $D$ is as in Theorem \ref{subfield_main}(\ref{item_2.2}) then $\tau(C_{D})$ is a binary $[128, 5, 64]$-linear code with the Hamming weight enumerator $X^{128} + 30 X^{64}Y^{64} + Y^{128}$. Note that $\tau(C_{D})$ is optimal according to the Database \cite{BoundTable}. It is self-orthogonal.
         		      \end{enumerate}
         	   \item Let $m=4$, $L=\{1, 2, 3\}$, $M=\{1, 2\}$, $N=\{4\}$. If $D$ is as in Theorem \ref{subfield_main}(\ref{item_3}) then $\tau(C_{D})$ is a binary $[192, 3, 96]$-linear code with the Hamming weight enumerator $X^{192} + 7 X^{96}Y^{96}$. It is minimal and self-orthogonal.
         	   \item Let $m=4$, $L=\{1, 2, 3, 4\}$, $M=\{1, 2\}$, $N=\{3, 4\}$. If $D$ is as in Theorem \ref{subfield_main}(\ref{item_4}) then $\tau(C_{D})$ is a binary $[768, 4, 384]$-linear code with the Hamming weight enumerator $X^{768} + 15 X^{384}Y^{384}$. It is minimal and self-orthogonal.
         	   \item Let $m=4$, $L=\{1, 2\}$, $M=\{3\}$, $N=\{3, 4\}$. If $D$ is as in Theorem \ref{subfield_main}(\ref{item_5}) then $\tau(C_{D})$ is a binary $[672, 4, 336]$-linear code with the Hamming weight enumerator $X^{672} + 12 X^{336}Y^{336} + 3X^{224}Y^{448}$. It is minimal and self-orthogonal.
         	   \item Let $m=4$, $L=\{2, 3\}$, $M=\{2\}$, $N=\{4\}$. If $D$ is as in Theorem \ref{subfield_main}(\ref{item_6}) then $\tau(C_{D})$ is a binary $[336, 4, 168]$-linear code with the Hamming weight enumerator $X^{336} + 12 X^{168}Y^{168} + 3X^{112}Y^{224}$. It is minimal and self-orthogonal.
         	   \item Let $m=4$, $L=\{2, 3, 4\}$, $M=\{1, 2, 4\}$, $N=\{1, 2, 3\}$. If $D$ is as in Theorem \ref{subfield_main}(\ref{item_7}) then $\tau(C_{D})$ is a binary $[512, 3, 256]$-linear code with the Hamming weight enumerator $X^{512} + 7 X^{256}Y^{256}$. It is minimal and self-orthogonal.
         	   \item Let $m=4$, $L=\{2, 4\}$, $M=\{1, 2, 4\}$, $N=\{2, 3, 4\}$. If $D$ is as in Theorem \ref{subfield_main}(\ref{item_8}) then $\tau(C_{D})$ is a binary $[768, 4, 384]$-linear code with the Hamming weight enumerator $X^{768} + 12 X^{384}Y^{384}+3X^{256}Y^{512}$. It is minimal and self-orthogonal.
         	\end{enumerate}
         \end{example}
        \begin{remark}
        	By Theorem \ref{Bonisoli} all the binary $1$-weight codes of this manuscript are simplex codes and hence they are distance optimal.
        \end{remark}
        Recall, for any integer $t \geq 1$, the $1st$ order Reed-Muller code $\mathcal{R}(1, t)$ of length $2^t$ is a binary $[2^{t}, t+1, 2^{t-1}]$-linear code, which has each codeword of Hamming weight $2^{t-1}$ except the $\textbf{0}$ and $\textbf{1}$ codewords (see [\cite{FirstCourse}, Proposition 6.2.3]).\\
        Now, we have the following result.
        \begin{theorem}
        	Let $\tau(C_{D})$ be the binary code as in Theorem \ref{subfield_main}{\eqref{item_2}}. Set $\vert L \vert = m-1$ and $M = N = \emptyset $. Then, the parameters and the weight distributions of $\tau(C_{D})$ and $\mathcal{R}(1, m-1)$, the 1st order binary Reed-Muller code,  are same.
        	
        \end{theorem}
        
        \begin{example}
        	\begin{enumerate}
        		\item Set $m=5$, $L=\{1, 2, 3, 4\}$, $M=N=\emptyset$ in Theorem \ref{subfield_main}{\eqref{item_2}}. Then $\tau(C_{D})$ is the 1st order binary Reed-Muller code $\mathcal{R}(1, 4)$ with parameters $[16, 5, 8]$.
        		\item  Set $m=6$, $L=\{ 2, 3, 4, 5, 6\}$, $M=N=\emptyset$ in Theorem \ref{subfield_main}{\eqref{item_2}}. Then $\tau(C_{D})$ is the 1st order binary Reed-Muller code $\mathcal{R}(1, 5)$ with parameters $[32, 6, 16]$.
        		\item Set $m=7$, $L=\{1, 2, 3, 4, 6, 7\}$, $M=N=\emptyset$ in Theorem \ref{subfield_main}{\eqref{item_2}}. Then $\tau(C_{D})$ is the 1st order binary Reed-Muller code $\mathcal{R}(1, 6)$ with $[64, 7, 32]$.
        	\end{enumerate}
        \end{example}
         
           \subsection{Using simplicial complexes having two maximal elements}
           This subsection studies the subfield-like code $\tau(C_D)$ where $\tau$ is defined by Equation \eqref{tauMap}, for various choices of $D$ defined by simplicial complexes with two maximal elements.
           Let $\Delta$ be a simplicial complex with two maximal elements $L$ and $M$. Now, we have the following result.
           \begin{theorem}\label{2subfield_main}
           	Let $\tau$ be the map defined by Equation \eqref{tauMap}. Suppose $m\in \mathbb{N}$ and $L, M \subseteq [m]$.
           	\begin{enumerate}
           		\item \label{2item_1}
           		Let $D = (1+u^2)\Delta + u^2\Delta + (u+u^2)\Delta \subseteq \mathcal{R}^m$. Then $\tau(C_D)$ is a binary $[(2^{\vert L \vert } + 2^{\vert M \vert } -2^{\vert L \cap M\vert })^3, \vert L\cup M \vert, 2^{\text{min}\{ \vert L \vert,  \vert M \vert \}-1}\times (2^{\vert L \vert } + 2^{\vert M \vert } -2^{\vert L \cap M \vert })^{2}]$ linear $4$-weight code and its weight distribution is displayed in Table \ref{2subfield_1}. In addition, it is self-orthogonal provided $\vert L \cap M\vert \geq 1$.
           		\begin{table}[]
           			\centering
           			\scalebox{0.7}{\begin{tabular}{  c | c  }
           				\hline
           				Hamming weight   & Number of codewords \\
           				\hline
           				$\frac{1}{2}(2^{\vert L \vert} + 2^{\vert M \vert} - 2^{\vert L \cap M\vert})^3$ & $2^{\vert L \cup M \vert} -2^{\vert L\setminus M \vert + \vert M\setminus L \vert}$\\        				
           				\hline
           				$\frac{1}{2}(2^{\vert L \vert} + 2^{\vert M \vert} - 2^{\vert L \cap M\vert})^2 (2^{\vert L \vert} + 2^{\vert M \vert} )$ & $2^{\vert L\setminus M \vert + \vert M \setminus L \vert } -2^{\vert L \setminus M \vert } -2^{\vert M \setminus L \vert } +1$ \\
           				\hline
           				$2^{\vert L \vert -1}(2^{\vert L \vert} + 2^{\vert M \vert} - 2^{\vert L \cap M\vert})^2  $ & $2^{\vert L\setminus M \vert } - 1$ \\
           				\hline
           				$2^{\vert M \vert -1}(2^{\vert L \vert} + 2^{\vert M \vert} - 2^{\vert L \cap M\vert})^2  $ & $2^{\vert M \setminus L \vert } - 1$ \\
           				\hline
           				$0$ & $1$\\
           				\hline
           			\end{tabular}}
           			\caption{Hamming weight distribution in Theorem \ref{2subfield_main}(\ref{2item_1})}
           			\label{2subfield_1}	
           		\end{table}
           		\item \label{2item_2}
           		Let $D = (1+u^2)\Delta^{\textnormal{c}} + u^2\Delta + (u+u^2)\Delta \subseteq \mathcal{R}^m$. Then $\tau(C_D)$ is a binary $[(2^{\vert L \vert } + 2^{\vert M \vert } -2^{\vert L \cap M\vert })^2(2^m- 2^{\vert L \vert } - 2^{\vert M \vert } +2^{\vert L \cap M\vert }), m, (2^{\vert L \vert } + 2^{\vert M \vert} -2^{\vert L \cap M \vert})^2 (2^{m-1} -2^{\vert L\vert -1} -2^{\vert M \vert -1} + 2^{\vert L \cap M\vert -1})]$ linear $5$-weight code and its weight distribution is displayed in Table \ref{2subfield_2}. If $\text{min}\{ \vert L \vert, \vert M \vert \} \leq m-3$ and $\text{max}\{ \vert L \vert, \vert M \vert \} \leq m-2$ then $\tau(C_D)$ is minimal. In addition, it is self-orthogonal provided $\vert L \cap M \vert  \geq 1$.
           		\begin{table}[]
           			\centering
           			\scalebox{0.7}{\begin{tabular}{  c | c  }
           					\hline
           					Hamming weight   & Number of codewords \\
           					\hline
           					$2^{m -1}(2^{\vert L \vert} + 2^{\vert M \vert} -2^{\vert L \cap M \vert})^2$ & $2^{m-\vert L \cup M \vert} -1$\\
           					\hline
           					$(2^{m -1} - 2^{\vert M \vert -1})(2^{\vert L \vert} + 2^{\vert M \vert} -2^{\vert L \cap M \vert})^2$ & $2^{m-\vert L \cup M \vert} (2^{\vert M\setminus L \vert} -1)$\\       				
           					\hline
           					$(2^{m -1} - 2^{\vert L \vert -1})(2^{\vert L \vert} + 2^{\vert M \vert} -2^{\vert L \cap M \vert})^2$ & $2^{m-\vert L \cup M \vert} (2^{\vert L \setminus M \vert} -1)$\\       				
           					\hline
           					$(2^{m -1} - 2^{\vert L \vert -1} - 2^{\vert M \vert -1} + 2^{\vert L\cap M \vert -1})(2^{\vert L \vert} + 2^{\vert M \vert} -2^{\vert L \cap M \vert})^2$ & $2^{m-\vert L \cap M \vert} (2^{\vert L\cap M \vert} -1)$\\       				
           					\hline
           					$(2^{m -1} - 2^{\vert L \vert -1} - 2^{\vert M \vert -1})(2^{\vert L \vert} + 2^{\vert M \vert} -2^{\vert L \cap M \vert})^2$ & $2^{m-\vert L \cup M \vert} (2^{\vert L\setminus M \vert} -1)(2^{\vert M\setminus L \vert} -1)$\\       				
           					\hline
           					$0$ & $1$\\
           					\hline
           			\end{tabular}}
           			\caption{Hamming weight distribution in Theorem \ref{2subfield_main}(\ref{2item_2})}
           			\label{2subfield_2}	
           		\end{table}
           		\item \label{2item_3}
           		Let $D = (1+u^2)\Delta+ u^2\Delta^{\textnormal{c}} + (u+u^2)\Delta  \text{ or }  (1+u^2)\Delta+ u^2\Delta + (u+u^2)\Delta^{\textnormal{c}} \subseteq \mathcal{R}^m$. Then $\tau(C_D)$ is a binary $[(2^{\vert L \vert } + 2^{\vert M \vert} - 2^{\vert L \cap M\vert })^{2} (2^m - 2^{\vert L \vert} - 2^{\vert M \vert} + 2^{\vert L \cap M \vert}), \vert L\cup M \vert, 2^{\text{min}\{ \vert L \vert,  \vert M \vert \}-1}\times (2^{\vert L \vert } + 2^{\vert M \vert} - 2^{\vert L \cap M\vert }) (2^m - 2^{\vert L \vert} - 2^{\vert M \vert} + 2^{\vert L \cap M \vert})]$ linear $4$-weight code and its weight distribution is displayed in Table \ref{2subfield_3}. In addition, it is self-orthogonal provided $\vert L \cap M \vert  \geq 1$.
           		\begin{table}[]
           			\centering
           			\scalebox{0.65}{\begin{tabular}{  c | c  }
           					\hline
           					Hamming weight   & Number of codewords \\
           					\hline
           					$(2^{\vert L \vert } +2^{\vert M \vert} -2^{\vert L \cap M\vert})(2^m - 2^{\vert L \vert } -2^{\vert M \vert} +2^{\vert L \cap M\vert})2^{\vert L \vert -1}$ & $2^{\vert L\setminus M \vert} -1$\\        				
           					\hline
           					$(2^{\vert L \vert } +2^{\vert M \vert} -2^{\vert L \cap M\vert})(2^m - 2^{\vert L \vert } -2^{\vert M \vert} +2^{\vert L \cap M\vert})2^{\vert M \vert -1}$ & $2^{\vert M\setminus L \vert} -1$\\        				
           					\hline
           					$(2^{\vert L \vert } +2^{\vert M \vert} -2^{\vert L \cap M\vert})(2^m - 2^{\vert L \vert } -2^{\vert M \vert} +2^{\vert L \cap M\vert})(2^{\vert L \vert -1} + 2^{\vert M \vert -1})$ & $(2^{\vert L\setminus M \vert} -1) (2^{\vert M\setminus L \vert} -1)$\\        				
           					\hline
           					$(2^{\vert L \vert } +2^{\vert M \vert} -2^{\vert L \cap M\vert})(2^m - 2^{\vert L \vert } -2^{\vert M \vert} +2^{\vert L \cap M\vert})\times $ & $(2^{\vert L\cup M \vert} - 2^{\vert L\setminus M \vert + \vert M\setminus L \vert} )$\\
           					$(2^{\vert L \vert -1} + 2^{\vert M \vert -1} -2^{\vert L \cap M\vert -1})$ & \\		
           					\hline
           					
           					$0$ & $1$\\
           					\hline
           			\end{tabular}}
           			\caption{Hamming weight distribution in Theorem \ref{2subfield_main}(\ref{2item_3})}
           			\label{2subfield_3}	
           		\end{table}
           		\item \label{2item_4}
           		Let $D = (1+u^2)\Delta^{\textnormal{c}} + u^2\Delta^{\textnormal{c}} + (u+u^2)\Delta \text{ or }(1+u^2)\Delta^{\textnormal{c}} + u^2\Delta + (u+u^2)\Delta^{\textnormal{c}} \subseteq \mathcal{R}^m$. Then $\tau(C_D)$ is a binary $[(2^{\vert L \vert} + 2^{\vert M \vert} -2^{\vert L \cap M \vert})(2^m - 2^{\vert L \vert} - 2^{\vert M \vert} + 2^{\vert L \cap M \vert})^2, m, (2^{m -1} - 2^{\vert L \vert -1} - 2^{\vert M \vert -1})(2^{\vert L \vert} + 2^{\vert M \vert} -2^{\vert L \cap M \vert})(2^m - 2^{\vert L \vert} - 2^{\vert M \vert} + 2^{\vert L \cap M \vert})]$ linear $5$-weight code and its weight distribution is displayed in Table \ref{2subfield_4}. If $\text{min}\{ \vert L \vert, \vert M \vert \} \leq m-3$ and $\text{max}\{ \vert L \vert, \vert M \vert \} \leq m-2$ then it is minimal. In addition, it is self-orthogonal provided $\vert L \cap M \vert  \geq 1$.
           		\begin{table}[]
           			\centering
           			\scalebox{0.6}{\begin{tabular}{  c | c  }
           					\hline
           					Hamming weight   & Number of codewords \\
           					\hline
           					$2^{m -1}(2^{\vert L \vert} + 2^{\vert M \vert} -2^{\vert L \cap M \vert})(2^m - 2^{\vert L \vert} - 2^{\vert M \vert} + 2^{\vert L \cap M \vert})$ & $2^{m-\vert L \cup M \vert} -1$\\
           					\hline
           					$(2^{m -1} - 2^{\vert M \vert -1})(2^{\vert L \vert} + 2^{\vert M \vert} -2^{\vert L \cap M \vert})(2^m - 2^{\vert L \vert} - 2^{\vert M \vert} + 2^{\vert L \cap M \vert})$ & $2^{m-\vert L \cup M \vert} (2^{\vert M\setminus L \vert} -1)$\\       				
           					\hline
           					$(2^{m -1} - 2^{\vert L \vert -1})(2^{\vert L \vert} + 2^{\vert M \vert} -2^{\vert L \cap M \vert})(2^m - 2^{\vert L \vert} - 2^{\vert M \vert} + 2^{\vert L \cap M \vert})$ & $2^{m-\vert L \cup M \vert} (2^{\vert L \setminus M \vert} -1)$\\       				
           					\hline
           					$(2^{m -1} - 2^{\vert L \vert -1} - 2^{\vert M \vert -1} + 2^{\vert L\cap M \vert -1})(2^{\vert L \vert} + 2^{\vert M \vert} -2^{\vert L \cap M \vert})\times $ & $2^{m-\vert L \cap M \vert} (2^{\vert L\cap M \vert} -1)$\\
           					$(2^m - 2^{\vert L \vert} - 2^{\vert M \vert} + 2^{\vert L \cap M \vert})$ & \\		
           					\hline
           					$(2^{m -1} - 2^{\vert L \vert -1} - 2^{\vert M \vert -1})(2^{\vert L \vert} + 2^{\vert M \vert} -2^{\vert L \cap M \vert})\times $ & $2^{m-\vert L \cup M \vert} (2^{\vert L\setminus M \vert} -1)(2^{\vert M\setminus L \vert} -1)$\\
           					$(2^m - 2^{\vert L \vert} - 2^{\vert M \vert} + 2^{\vert L \cap M \vert})$ & \\  				
           					\hline
           					$0$ & $1$\\
           					\hline
           			\end{tabular}}
           			\caption{Hamming weight distribution in Theorem \ref{2subfield_main}(\ref{2item_4})}
           			\label{2subfield_4}	
           		\end{table}
           		\item \label{2item_5}
           		Let $D = (1+u^2)\Delta + u^2\Delta^{\textnormal{c}} + (u+u^2)\Delta^{\textnormal{c}} \subseteq \mathcal{R}^m$. Then $\tau(C_D)$ is a binary $[(2^{\vert L 
           			vert} +2^{\vert M \vert } -2^{\vert L \cap M \vert })(2^m - 2^{\vert L \vert} - 2^{\vert M \vert} + 2^{\vert L \cap M\vert })^2, \vert L \cup M\vert , 2^{\text{min}\{ \vert L \vert,  \vert M \vert \}-1}\times (2^m - 2^{\vert L \vert} - 2^{\vert M \vert} + 2^{\vert L \cap M \vert})^2]$ linear $4$-weight code and its weight distribution is displayed in Table \ref{2subfield_5}. In addition, it is self-orthogonal provided $\vert L \cap M\vert  \geq 1$.
           		\begin{table}[]
           			\centering
           			\scalebox{0.7}{\begin{tabular}{  c | c  }
           					\hline
           					Hamming weight   & Number of codewords \\
           					\hline
           					$2^{\vert L \vert -1}(2^m - 2^{\vert L \vert } -2^{\vert M \vert} +2^{\vert L \cap M\vert})^2 $ & $2^{\vert L\setminus M \vert} -1$\\        				
           					\hline
           					$2^{\vert M \vert -1} (2^m - 2^{\vert L \vert } -2^{\vert M \vert} +2^{\vert L \cap M\vert})^2 $ & $2^{\vert M\setminus L \vert} -1$\\        				
           					\hline
           					$(2^{\vert L \vert -1} + 2^{\vert M \vert -1})(2^m - 2^{\vert L \vert } -2^{\vert M \vert} +2^{\vert L \cap M\vert})^2$ & $(2^{\vert L\setminus M \vert} -1) (2^{\vert M\setminus L \vert} -1)$\\        				
           					\hline
           					$(2^{\vert L \vert -1} + 2^{\vert M \vert -1} -2^{\vert L \cap M\vert -1})(2^m - 2^{\vert L \vert } -2^{\vert M \vert} +2^{\vert L \cap M\vert})^2$ & $(2^{\vert L\cup M \vert} - 2^{\vert L\setminus M \vert + \vert M\setminus L \vert} )$\\        				
           					\hline
           					$0$ & $1$\\
           					\hline
           			\end{tabular}}
           			\caption{Hamming weight distribution in Theorem \ref{2subfield_main}(\ref{2item_5})}
           			\label{2subfield_5}	
           		\end{table}
           		\item \label{2item_6}
           		Let $D = (1+u^2)\Delta^{\textnormal{c}} + u^2\Delta^{\textnormal{c}} + (u+u^2)\Delta_N^{\textnormal{c}} \subseteq \mathcal{R}^m$. Then $\tau(C_D)$ is a binary $[(2^m - 2^{\vert L \vert} - 2^{\vert M \vert} + 2^{\vert L \cap M \vert})^3, m, (2^{m -1} - 2^{\vert L \vert -1} - 2^{\vert M \vert -1})(2^m - 2^{\vert L \vert} - 2^{\vert M \vert} + 2^{\vert L \cap M \vert})^2]$ linear $5$-weight code and its weight distribution is displayed in Table \ref{2subfield_6}. If $\text{min}\{ \vert L \vert, \vert M \vert \} \leq m-3$ and $\text{max}\{ \vert L \vert, \vert M \vert \} \leq m-2$ then it is minimal. In addition, it is self-orthogonal provided $\vert L \cap M \vert  \geq 1$.
           		\begin{table}[]
           			\centering
           			\scalebox{0.7}{\begin{tabular}{  c | c  }
           					\hline
           					Hamming weight   & Number of codewords \\
           					\hline
           					$2^{m -1}(2^m - 2^{\vert L \vert} - 2^{\vert M \vert} + 2^{\vert L \cap M \vert})^2$ & $2^{m-\vert L \cup M \vert} -1$\\
           					\hline
           					$(2^{m -1} - 2^{\vert M \vert -1})(2^m - 2^{\vert L \vert} - 2^{\vert M \vert} + 2^{\vert L \cap M \vert})^2$ & $2^{m-\vert L \cup M \vert} (2^{\vert M\setminus L \vert} -1)$\\       				
           					\hline
           					$(2^{m -1} - 2^{\vert L \vert -1})(2^m - 2^{\vert L \vert} - 2^{\vert M \vert} + 2^{\vert L \cap M \vert})^2$ & $2^{m-\vert L \cup M \vert} (2^{\vert L \setminus M \vert} -1)$\\       				
           					\hline
           					$(2^{m -1} - 2^{\vert L \vert -1} - 2^{\vert M \vert -1} + 2^{\vert L\cap M \vert -1})(2^m - 2^{\vert L \vert} - 2^{\vert M \vert} + 2^{\vert L \cap M \vert})^2$ & $2^{m-\vert L \cap M \vert} (2^{\vert L\cap M \vert} -1)$\\       				
           					\hline
           					$(2^{m -1} - 2^{\vert L \vert -1} - 2^{\vert M \vert -1})(2^m - 2^{\vert L \vert} - 2^{\vert M \vert} + 2^{\vert L \cap M \vert})^2$ & $2^{m-\vert L \cup M \vert} (2^{\vert L\setminus M \vert} -1)(2^{\vert M\setminus L \vert} -1)$\\       				
           					\hline
           					$0$ & $1$\\
           					\hline
           			\end{tabular}}
           			\caption{Hamming weight distribution in Theorem \ref{2subfield_main}(\ref{2item_6})}
           			\label{2subfield_6}	
           		\end{table}
           	\end{enumerate}        	
           \end{theorem}
           \proof We prove part \eqref{2item_2} and other parts can be proved similarly.\\
           Note that the length of $\tau(C_{D})$ is $\vert D \vert = (2^m - 2^{\vert L \vert} - 2^{\vert M \vert} +2^{\vert L \cap M \vert})(2^{\vert L \vert} + 2^{\vert M \vert} -2^{\vert L \cap M \vert})^2$.
           Suppose $x = (1+u^2)\alpha + u^2 \beta + (u+u^2)\gamma \in \mathcal{R}^m $ and $D = (1+u^2)\Delta^{\textnormal{c}} + u^2\Delta + (u + u^2)\Delta $. Then by Lemma \ref{Complement_sum}, Equation \eqref{DefinePsiFunction} and Equation \eqref{KeyEq_SF_like}, we have
           \begin{equation*}
           	\begin{split}
           		wt_{H}\big(\tau(c_D(x))\big)  
           		  = & \frac{1}{2}(2^{\vert L \vert} + 2^{\vert M \vert} -2^{\vert L \cap M \vert})^2 \big[(2^m - 2^{\vert L \vert} - 2^{\vert M \vert} +2^{\vert L \cap M \vert}) - (2^m \delta_{0, \alpha} - \Psi_{\alpha}(\Delta))\big].
           	\end{split}
           \end{equation*}
           Here we discuss the following cases.
           \begin{enumerate}
           	\item If $\alpha =0$ then $wt_{H}\big(\tau(c_D(x))\big) =0$.\\
           	In this case, $\# \alpha =1 , \# \beta =2^m, \# \gamma =2^m$. Therefore, $Z_0 = 2^{2m}$.
           	\item If $\alpha \neq 0$ then $wt_{H}\big(\tau(c_D(x))\big) = \frac{1}{2}(2^{\vert L \vert} + 2^{\vert M \vert} -2^{\vert L \cap M \vert})^2 \big[(2^m - 2^{\vert L \vert} - 2^{\vert M \vert} +2^{\vert L \cap M \vert}) + \Psi_{\alpha}(\Delta)\big]$.
           	\begin{itemize}
           		\item If $\alpha \in \mathfrak{U}_1$ then $wt_{H}\big(\tau(c_D(x))\big) = 2^{m-1}(2^{\vert L \vert} + 2^{\vert M \vert} -2^{\vert L \cap M \vert})^2$.\\
           		In this case, we get\\
           		$\# \alpha = 2^{m- \vert L \cup M \vert }-1$,
           		$\# \beta = 2^m$,
           		$\# \gamma =2^m$.\\
           		Therefore, $\#x = (2^{m- \vert L \cup M \vert }-1)\times 2^{2m}$.
           		
           		\item If $\alpha \in \mathfrak{U}_2$ then $wt_{H}\big(\tau(c_D(x))\big) = (2^{m-1} - 2^{\vert M \vert -1})(2^{\vert L \vert} + 2^{\vert M \vert} -2^{\vert L \cap M \vert})^2$.\\
           		In this case, we get\\
           		$\# \alpha = (2^{\vert M\setminus L \vert } -1)2^{m- \vert L \cup M \vert }$,
           		$\# \beta = 2^m$,
           		$\# \gamma =2^m$.\\
           		Therefore, $\#x = (2^{\vert M\setminus L \vert } -1)2^{3m- \vert L \cup M \vert }$.
           		
           		\item If $\alpha \in \mathfrak{U}_3$ then $wt_{H}\big(\tau(c_D(x))\big) = (2^{m-1} - 2^{\vert L \vert -1})(2^{\vert L \vert} + 2^{\vert M \vert} -2^{\vert L \cap M \vert})^2$.\\
           		In this case, we get\\
           		$\# \alpha = (2^{\vert L\setminus M \vert } -1)2^{m- \vert L \cup M \vert }$,
           		$\# \beta = 2^m$,
           		$\# \gamma =2^m$.\\
           		Therefore, $\#x = (2^{\vert L\setminus M \vert } -1)2^{3m- \vert L \cup M \vert }$.
           		
           		\item If $\alpha \in \mathfrak{U}_4$ then $wt_{H}\big(\tau(c_D(x))\big) = (2^{m-1} - 2^{\vert L \vert -1}- 2^{\vert M \vert -1})(2^{\vert L \vert} + 2^{\vert M \vert} -2^{\vert L \cap M \vert})^2$.\\
           		In this case, we get\\
           		$\# \alpha = (2^{\vert L\setminus M \vert } -1)(2^{\vert M\setminus L \vert } -1)2^{m- \vert L \cup M \vert }$,
           		$\# \beta = 2^m$,
           		$\# \gamma =2^m$.\\
           		Therefore, $\#x = (2^{\vert L\setminus M \vert } -1)(2^{\vert M\setminus L \vert } -1)2^{3m- \vert L \cup M \vert }$.
           		
           		\item If $\alpha \in \mathfrak{U}_5$ then $wt_{H}\big(\tau(c_D(x))\big) = (2^{m-1} - 2^{\vert L \vert -1}- 2^{\vert M \vert -1} + 2^{\vert L \cap M \vert -1})(2^{\vert L \vert} + 2^{\vert M \vert} -2^{\vert L \cap M \vert})^2$.\\
           		In this case, we get\\
           		$\# \alpha = (2^{\vert L\cap M \vert } -1)2^{m- \vert L \cap M \vert }$,
           		$\# \beta = 2^m$,
           		$\# \gamma =2^m$.\\
           		Therefore, $\#x = (2^{\vert L\cap M \vert } -1)2^{3m- \vert L \cap M \vert }$.
           	\end{itemize}
           \end{enumerate}
           Note that $T: \mathcal{R}^m \longrightarrow \tau(C_{D}) \subseteq \mathbb{F}_{2}^{\vert D\vert }$ is a surjective homomorphism and $\vert \ker(T) \vert = \vert \{x \in \mathcal{R}^m : \tau(x\cdot d) = 0 ~ \forall ~d \in D\} \vert  = 2^{2m}$. By the first isomorphism theorem, we have $\vert \tau(C_{D}) \vert = \frac{\vert \mathcal{R}^m \vert}{\vert \ker(T) \vert} = 2^m$. Hence, $\dim(\tau(C_{D})) = m$.
           
           \begin{table}[h!]
           	\centering      		
           	\scalebox{0.7}{\begin{tabular}{c|c|c|c|c || c|c|c|c|c}
           			\hline
           			S.N. & Result & \#Weight & Optimal  & Minimal & S.N. & Result & \#Weight & Optimal  & Minimal \\
           			\hline
           			$1$ & Theorem \ref{sec4theorem} (1)  &  $2$ & $-$ & Yes & $11$ & Theorem \ref{subfield_main} (\ref{item_3})& 1 & Yes & Yes\\
           			\hline
           			$2$ & Theorem \ref{sec4theorem} (2)  &  $4$  & $-$ & Yes & $12$ & Theorem \ref{subfield_main} (\ref{item_4})& 1 & Yes & Yes\\ 
           			\hline
           			$3$ & Theorem \ref{sec4theorem} (3)  &  5    & $-$ & Yes & $13$ & Theorem \ref{subfield_main} (\ref{item_5})& 2 & $-$ &  Yes\\
           			\hline
           			$4$ & Theorem \ref{sec4theorem} (4)  & 4    & $-$ & Yes & $14$ & Theorem \ref{subfield_main} (\ref{item_6})& 2 & $-$ & Yes\\
           			\hline
           			$5$ & Theorem \ref{sec4theorem} (5)  & 10   & $-$ & Yes & $15$ & Theorem \ref{subfield_main} (\ref{item_7})& 1 & Yes & Yes\\
           			\hline
           			$6$ & Theorem \ref{sec4theorem} (6)  & 8    & $-$ & Yes & $16$ & Theorem \ref{subfield_main} (\ref{item_8})& 2 & $-$ & Yes\\
           			\hline
           			$7$ & Theorem \ref{sec4theorem} (7)  & 8    & $-$ & Yes & $17$ & Theorem \ref{2subfield_main} (\ref{2item_2})& 5 & $-$ & Yes\\
           			\hline
           			$8$ & Theorem \ref{sec4theorem} (8)  &  16    & $-$ & Yes & $18$ &  Theorem \ref{2subfield_main} (\ref{2item_4})& 5 & $-$ & Yes\\
           			\hline
           			$9$ & Theorem \ref{subfield_main} (\ref{item_1})& 1 & Yes & Yes & $19$ & Theorem \ref{2subfield_main} (\ref{2item_6})& 5 & $-$  & Yes\\
           			\hline
           			$10$ & Theorem \ref{subfield_main} (\ref{item_2})& 2 & Yes & Yes & & & & & \\
           			\hline
           	\end{tabular}}
           	\caption{Binary linear codes from simplicial complexes in this manuscript}
           	\label{table:article}        	
           \end{table}
           
           \begin{table}[h!]
           	\centering
           	\begin{adjustbox}{width=\textwidth}
           		
           		\begin{tabular}{|c|c|c|c|c|c|}
           			\hline
           			Reference  & Result   & $[n,k,d]$-code & \#Weight & Bound  & Minimal\\
           			\hline
           			\multirow{4}{*}{\cite{Wu_Li}}&  Theorem $5.2$&$[2^{2m}-2^{ \vert A \vert + \vert B \vert }, 2m, 2^{2m-1}-2^{ \vert A \vert + \vert B \vert -1}]$&$2$ &Griesmer & $-$\\
           			\cline{2-6}
           			&  Proposition $5.3$ &$[(2^{ \vert A \vert }+2^{ \vert B \vert }-2^{ \vert A\cap B \vert })^2-1, 2 \vert A\cup B \vert ]$&$\leq 10$&$-$ & $-$\\
           			\cline{2-6}                                            
           			&  Theorem $5.5$  &$[(2^{ \vert A \vert }+2^{ \vert B \vert }-2^{ \vert A\cap B \vert })^2-1, (2^{ \vert A \vert }+2^{ \vert B \vert }-2^{ \vert A\cap B \vert })^2-1-2 \vert A\cup B \vert , 3]$& &Sphere packing & $-$\\
           			\cline{2-6}
           			&  Proposition $5.6$ &$[4^m-(2^{ \vert A \vert }+2^{ \vert B \vert }-2^{ \vert A\cap B \vert })^2, 2m]$& $\leq 11$&  $-$  &$-$ \\              
           			\hline
           			
           			\multirow{2}{*}{\cite{Wu_Lee}}  &Theorem $5$  & $[2^{ \vert A \vert }-2^{ \vert B \vert },  \vert A \vert , 2^{ \vert A \vert -1}-2^{ \vert B \vert -1}]$ & 2 & Griesmer & $-$\\
           			\cline{2-6}
           			& Theorem $6$ &  $[2^{ \vert A \vert }-2^{ \vert B \vert }, 2^{ \vert A \vert }-2^{ \vert B \vert }- \vert A \vert , 3 \text{ or }4]$ &  & $-$ &$-$\\
           			\hline
           			
           			\cite{mixed1} & Theorem $4$ & $[(2^{m} - 2^{n})^3, 3m, 2^m(2^m - 2^n)(2^m-2^{n + 1})]$ & $9$  & $-$  & Yes, if $n < m-1 $\\
           			\hline
           			\multirow{4}{*}{\cite{Sagar_Sarma_I}} & Theorem $4.3 (2)(a) $ & \multirow{2}{*}{$ [(2^m - 2^{\vert M \vert})2^{\vert N \vert +1}, m, (2^m - 2^{\vert M \vert})2^{\vert N \vert }]$ } & \multirow{2}{*}{$2$} & Griesmer, if $1\leq \theta_1 < \vert M \vert + \vert N \vert +1 \leq m$ & \multirow{2}{*}{Yes, if $\vert M \vert \leq m-2$} \\
           			\cline{2-2}\cline{5-5}
           			& Theorem $4.3 (2)(b)$ & & & Griesmer, if $0 < \theta_2 < m \leq \vert M \vert + \vert N \vert $ & \\
           			\cline{2-6}
           			& Theorem $4.3 (4)$ & $ [2(2^m - 2^{\vert M \vert})(2^m - 2^{\vert N \vert}), \vert M \vert, (2^m - 2^{\vert M \vert})(2^m - 2^{\vert N \vert})] $ & $2$ & $-$ & Yes, if $\vert M \vert \leq m-2$\\
           			\cline{2-6}
           			& Theorem $4.3 (5)$ & $ [2(2^{2m} - 2^{\vert M \vert + \vert N \vert}),m, (2^{2m} - 2^{\vert M \vert + \vert N \vert})] $ & $2$ & $-$ & Yes, if $\vert M \vert + \vert N \vert \leq 2m-2$\\
           			\hline
           			\multirow{8}{*}{\cite{Sagar_Sarma_E}} & Theorem $4.2 (2a)$ &\multirow{2}{*}{$ [(2^m -2^{ \vert M \vert })2^{\vert N \vert }, m, (2^m - 2^{\vert M \vert })2^{\vert N \vert -1}] $} & \multirow{2}{*}{$2$} & Griesmer, if $0\leq \theta_1 < \vert M \vert + \vert N \vert \leq m $ & \multirow{2}{*}{Yes, if $\vert M \vert + \vert N \vert \geq 3$}\\
           			\cline{2-2} \cline{5-5}
           			 & Theorem $4.2 (2b)$ & & & Griesmer, if $0 < \theta_2 < m <\vert M \vert + \vert N \vert  $ & \\
           			\cline{2-6}
           			& Theorem $4.2 (4)$ & $ [(2^m - 2^{\vert M \vert})(2^m - 2^{\vert N \vert}), m, (2^m - 2^{\vert N \vert})(2^{m-1} - 2^{\vert N \vert -1})] $ & $2$ & $-$ & Yes, if $\vert M \vert \leq m-2 $\\
           			\cline{2-6}
           			& Theorem $4.2 (5)$ & $ [(2^{2m} - 2^{\vert M \vert + \vert N \vert}), m,(2^{2m -1} - 2^{\vert M \vert + \vert N \vert -1}) ] $ & $2$ & $-$ & Yes, if $\vert M \vert + \vert N \vert \leq 2m - 2$\\
           			\cline{2-6}
           			& Theorem $5.3 (3a)$ & \multirow{2}{*}{$ [2^{\vert M \vert +1}(2^m - 2^{\vert N \vert }), m, 2^{\vert M \vert }(2^m - 2^{\vert N \vert })] $} & \multirow{2}{*}{$3$} & Griesmer, if $1\leq \theta_3 < \vert M \vert + \vert N \vert +1 \leq m $ & \multirow{2}{*}{Yes, if $\vert N \vert \leq m-2$}\\
           			\cline{2-2} \cline{5-5}
           			& Theorem $5.3 (3b)$ & &  & Griesmer, if $0 < \theta_4 < m < \vert M \vert + \vert N \vert $ & \\
           			\cline{2-6}
           			& Theorem $5.3 (4)$ & $ [2(2^m - 2^{\vert M \vert })(2^m - 2^{\vert N \vert}), m, (2^m - 2^{\vert M \vert })(2^m - 2^{\vert N \vert})] $ & $3$ & $-$ & Yes\\
           			\cline{2-6}
           			& Theorem $5.3 (5)$ & $ [2(2^{2m} - 2^{\vert M \vert + \vert N \vert }), m, (2^{2m} - 2^{\vert M \vert + \vert N \vert })] $ & $3$ & $-$ & Yes, if $\vert M \vert + \vert N \vert \leq 2m -2$\\
           			\hline
           			\cite{Sagar_Sarma} &  Theorem $4.6$ & $[2^{3m}-2^{\vert L \vert + \vert M \vert + \vert N \vert }, 3m, 2^{3m-1} -2^{\vert L \vert + \vert M \vert + \vert N \vert -1}]$ & $2$ & Griesmer & Yes, if $\vert L \vert + \vert M \vert + \vert N \vert \leq 3m -2$\\
           			\hline
           			\multirow{8}{*}{\cite{Hyun_Lee}}  & Lemma $7$ &  $[2^{m}-2^{ \vert A \vert }, m, 2^{m-1}-2^{ \vert A \vert -1}]$ & 2 & Griesmer & $-$\\
           			\cline{2-6}
           			&                       Ex. $10$                 & $[2^{m-1}, m, 4]$ &  & Sphere Packing & $-$\\
           			\cline{2-6}
           			&                          Corollary $21$              & $[2^m-2\sum_{i=1}^{s}2^{ \vert A_i \vert -1}+s-1, m, 2^{m-1}-\sum_{i=1}^{s}2^{ \vert A_i \vert -1}]$ &  & Griesmer  & $-$\\
           			\cline{2-6}
           			&                    \multirow{2}{*}{Lemma $26$}  & $[2^{ \vert A_1 \vert }+2^{ \vert A_2 \vert }-2,  \vert A_1\cup A_2 \vert , 2^{ \vert A_1 \vert -1}]$ & 3 &  $-$&$-$ \\
           			\cline{3-6}
           			&                      & $[2^{ \vert A_1 \vert }+2^{ \vert A_2 \vert }-2^{ \vert A_1\cap A_2 \vert }-1,  \vert A_1\cup A_2 \vert , 2^{ \vert A_1 \vert -1}]$ & 4 & $-$&$-$ \\
           			\cline{2-6}
           			&                   \multirow{2}{*}{Theorem $27$}  & $[2^m-2^{ \vert A_1 \vert }-2^{ \vert A_2 \vert }+1, m, 2^{m-1}-2^{ \vert A_1 \vert -1}2^{ \vert A_2 \vert -1}]$ & 3 or 4 & Griesmer  & $-$\\
           			\cline{3-6}
           			&                    &   $[2^m-2^{ \vert A_1 \vert }-2^{ \vert A_2 \vert }+2^{ \vert A_1\cap A_2 \vert }-1, m, 2^{m-1}-2^{ \vert A_1 \vert -1}-2^{ \vert A_2 \vert -1}]$ & 4 or 5 & $-$&$-$ \\
           			\hline
           		\end{tabular}
           		
           	\end{adjustbox}
           	\caption{Binary linear codes from simplicial complexes before this manuscript}
           	\label{table:RecentComparision} 		
           \end{table}
           
           \begin{table}[h!]
           	\centering
           		
           		\scalebox{0.7}{\begin{tabular}{|c|c|c|c|c|c|}
           			\hline
           			$m$ &	$L$ & $M$  & $N$ & $[n,k,d]$-code & Remark   \\
           			\hline
           			\multirow{8}{*}{$m=5$}& $(1, 1, 0, 0, 0)$ & $(0, 0, 1, 0, 0)$  & $(0, 0, 0, 0, 0)$ & $[56, 5, 28]$& New\\
           			\cline{2-6}
           			&    $(0, 0, 1, 1, 0)$  & $(1, 0, 0, 0, 0)$    & $(0, 0, 0, 0, 1)$ &$[112, 5, 56]$ & New\\
           			\cline{2-6}
           			&     $(1, 1, 1, 0, 0)$   & $(0, 0, 0, 1, 0)$    & $(0, 0, 0, 0, 1)$ &$[96, 5, 48]$ & New\\
           			\cline{2-6}
           			&     $(1, 0, 0, 0, 0)$   & $(0, 0, 0, 0, 0)$    & $(0, 0, 0, 0, 0)$ &$[30, 5, 15]$ & $-$\\
           			\cline{2-6}
           			&       $(0, 1, 1, 1, 1)$   & $(0, 1, 0, 0, 0)$    & $(1, 0, 0, 0, 0)$&$[64, 5, 32]$ & New\\
           			\cline{2-6}
           			&       $(0, 1, 1, 1, 1)$   & $(0, 0, 1, 0, 0)$    & $(1, 1, 0, 0, 0)$&$[128, 5, 64]$ & New\\
           			\cline{2-6}
           			&       $(1, 1, 1, 1, 0)$   & $(0, 0, 0, 0, 0)$    & $(0, 0, 0, 0, 0)$&$[16, 5, 8]$ & $-$\\
           			\cline{2-6}
           			&       $(1, 1, 1, 0, 0)$   & $(1, 1, 0, 0, 0)$    & $(0, 0, 1, 0, 0)$&$[128, 5, 64]$ & New\\
           			\cline{1-6}
           			\multirow{8}{*}{$m=6$}
           			& $(1, 0, 0, 0, 0, 0)$ & $(0, 1, 0, 0, 0, 0)$  & $(0, 0, 0, 0, 0, 0)$ & $[124, 6, 62]$& New\\
           			\cline{2-6}
           			& $(1, 1, 0, 0, 0, 0)$ & $(0, 0, 1, 0, 0, 0)$  & $(0, 0, 0, 1, 0, 0)$ & $[240, 6, 120]$& New\\ 
           			\cline{2-6}
           			& $(1, 1, 1, 0, 0, 0)$ & $(0, 0, 0, 1, 0, 0)$  & $(0, 0, 0, 0, 1, 0)$ & $[224, 6, 112]$& New\\ 
           			\cline{2-6}
           			& $(0, 1, 1, 1, 1, 0)$ & $(0, 0, 0, 0, 0, 0)$  & $(0, 0, 0, 0, 0, 0)$ & $[48, 6, 24]$& $-$\\
           			\cline{2-6}
           			& $(1, 1, 1, 1, 1, 0)$ & $(1, 0, 0, 0, 0, 0)$  & $(0, 0, 0, 0, 0, 1)$ & $[128, 6, 64]$& New\\ 
           			\cline{2-6}
           			& $(1, 1, 1, 1, 1, 0)$ & $(1, 0, 0, 0, 0, 0)$  & $(0, 1, 1, 0, 0, 0)$ & $[256, 6, 128]$& New\\
           			\cline{2-6}
           			& $(1, 1, 1, 1, 1, 0)$ & $(1, 1, 0, 0, 0, 0)$  & $(0, 0, 0, 0, 0, 0)$ & $[128, 6, 64]$& New\\ 
           			\cline{2-6}
           			& $(0, 1, 1, 1, 1, 1)$ & $(0, 0, 0, 0, 0, 0)$  & $(0, 0, 0, 0, 0, 0)$ & $[32, 6, 16]$& $-$\\
           			\cline{1-6}
           			\multirow{8}{*}{$m=7$} 
           			& $(1, 0, 0, 0, 0, 0, 0)$ & $(0, 1, 0, 0, 0, 0, 0)$  & $(0, 0, 0, 0, 0, 0, 0)$ & $[252, 7, 126]$& New\\ 
           			\cline{2-6}
           			& $(1, 1, 0, 0, 0, 0, 0)$ & $(0, 0, 1, 0, 0, 0, 0)$  & $(0, 0, 0, 0, 0, 0, 0)$ & $[248, 7, 124]$& New\\ 
           			\cline{2-6}
           			& $(1, 1, 1, 0, 0, 0, 0)$ & $(0, 0, 0, 0, 0, 0, 0)$  & $(0, 0, 0, 1, 0, 0, 0)$ & $[240, 7, 120]$& New\\ 
           			\cline{2-6}
           			& $(0, 1, 1, 1, 1, 0, 0)$ & $(0, 0, 0, 0, 0, 0, 0)$  & $(0, 0, 0, 0, 0, 0, 0)$ & $[112, 7, 56]$& $-$\\ 
           			\cline{2-6}
           			& $(1, 1, 1, 1, 1, 1, 0)$ & $(0, 0, 0, 0, 0, 0, 0)$  & $(1, 1, 0, 0, 0, 0, 0)$ & $[256, 7, 128]$& New\\ 
           			\cline{2-6}
           			& $(1, 1, 1, 1, 1, 1, 0)$ & $(0, 0, 0, 0, 0, 0, 0)$  & $(1, 1, 0, 0, 0, 0, 0)$ & $[256, 7, 128]$& New\\
           			\cline{2-6}
           			& $(0, 1, 1, 1, 1, 1, 1)$ & $(0, 0, 0, 0, 0, 0, 0)$  & $(0, 0, 0, 0, 0, 0, 0)$ & $[64, 7, 32]$& $-$\\
           			\hline
           		\end{tabular}}
           	\caption{Optimal linear codes from Theorem \ref{subfield_main} (\ref{item_2})}
           	\label{table:ListNewCodes} 		
           \end{table}

        \section{Code comparison}\label{section6}
        In this section, we give three tables that demonstrate how our codes have a considerable benefit. For the reader's convenience, we list certain few-weight codes of this manuscript having good parameters in Table \ref{table:article}. Table \ref{table:RecentComparision} presents some recent works on linear codes constructed with the help of simplicial complexes.\\
        In Table \ref{table:ListNewCodes}, we present several binary codes which are obtained from Theorem \ref{subfield_main} (\ref{item_2}). We verified with MAGMA that all the codes in this table are optimal. Remark ``New'' in the table indicates that the corresponding codes are new in the sense that they are inequivalent to the currently best-known linear codes. 
        
		\section{Conclusion}\label{section7}
		 In this manuscript, certain linear codes over the ring $\mathbb{F}_2[u]/\langle u^3-u \rangle$ are studied with the help of simplicial complexes generated by one and two maximal elements. Under very mild conditions, their Gray images and their binary subfield-like codes corresponding to a certain $\mathbb{F}_{2}$-functional of $\mathcal{R}$, are minimal codes. Most of the binary codes here are few-weight codes. The binary codes in this manuscript are self-orthogonal under certain sufficient conditions. Besides, we obtain an infinite family of optimal codes with respect to the Griesmer bound.\\ 
		 $~~~~$ In future, one can study linear codes over the ring $\mathbb{F}_2[u]/\langle u^t-u \rangle$ for any $t \geq 4$ using simplicial complexes by devising an elegant method and obtain linear codes with nice parameters.\\~\\
		 \textbf{Acknowledgements } The authors thank Professor Yansheng Wu for helpful suggestions. The authors would also like to thank the handling editor and the anonymous referees for their valuable comments and suggestions which have highly improved the quality of the paper.

	\end{document}